%% file: main.tex
\documentclass[lettersize,journal]{IEEEtran}
\usepackage{amsmath,amsfonts}
\usepackage{algorithmic}
\usepackage{algorithm}
\usepackage{array}
\usepackage[caption=false,font=normalsize,labelfont=sf,textfont=sf]{subfig}
\usepackage{textcomp}
\usepackage{stfloats}
\usepackage{url}
\usepackage{verbatim}
\usepackage{graphicx}
\usepackage{cite}
\input{macros}

\begin{document}

\title{Quality assessment of 3D human animation: Subjective and objective evaluation}

\author{Rim Rekik$^1$ \and Stefanie Wuhrer$^1$ \and Ludovic Hoyet$^2$ \and Katja Zibrek$^2$ \and Anne-H\'{e}l\`{e}ne Olivier$^2$
\thanks{$^1$ Inria centre at the University Grenoble Alpes, France}
\thanks{$^2$ Inria, Univ Rennes, CNRS, IRISA, France}
}

\markboth{}%
{}


\maketitle
\begin{figure*}[h]
    \centering
    \includegraphics[width=\linewidth]{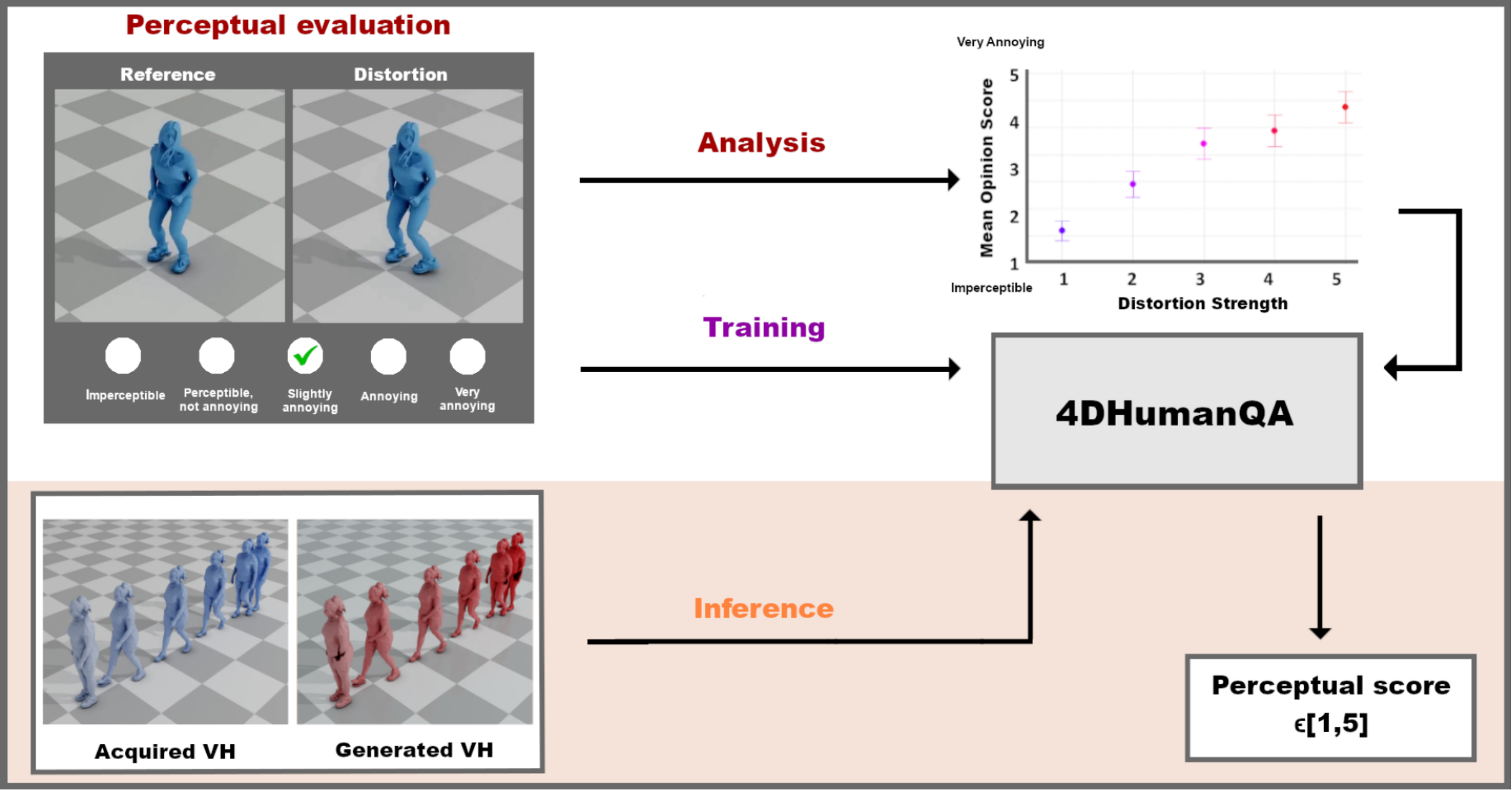}
    \caption{We conduct a perceptual evaluation to collect subjective scores for visual distortions of generated 3D human animations with respect to corresponding references, which are the acquired 3D reconstructions of real actors. We use the resulting \emph{``4DHumanPercept''} dataset to first analyse the factors influencing human motion realism, and second, to learn a data-driven model called \emph{``4DHumanQA''} that predicts a perceptual score for 3D human animation realism.
    }
    \label{fig:Teaser}
\end{figure*}

\begin{abstract}
\input{sections/abstract}
\end{abstract}

\begin{IEEEkeywords}
Computer graphics, perception, visual quality assessment, subjective quality evaluation, objective quality evaluation, dataset, perceptual metric, human animation, 3D digital human evaluation.
\end{IEEEkeywords}

\input{sections/intro}

\input{sections/related-work}
\input{sections/overview}

\input{sections/stimuli-generation}

\input{sections/subjective-exp}

\input{sections/metric}

\input{sections/conclusion}
\input{sections/Acknowledgments}
\bibliographystyle{abbrv}
\bibliography{main}
\input{supplementary/spl}

\end{document}

%% file: macros.tex
\usepackage{xcolor}
\usepackage[normalem]{ulem}

\newcommand{\Fval}[4][]{F#1$_{#2,#3}$=#4}

%% file: sections/abstract.tex
Virtual human animations have a wide range of applications in virtual and augmented reality. While automatic generation methods of animated virtual humans have been developed, assessing their quality remains challenging. Recently, approaches introducing task-oriented evaluation metrics have been proposed, leveraging neural network training. However, quality assessment measures for animated virtual humans that are not generated with parametric body models have yet to be developed. In this context, we introduce a first such quality assessment measure leveraging a novel data-driven framework. First, we generate a dataset of virtual human animations together with their corresponding subjective realism evaluation scores collected with a user study. Second, we use the resulting dataset to learn predicting perceptual evaluation scores. Results indicate that training a linear regressor on our dataset results in a correlation of $90\%$, which outperforms a state of the art deep learning baseline.

%% file: sections/intro.tex
\section{Introduction}

\IEEEPARstart{V}{irtual} Human (VH) animations have multiple applications~\cite{carrozzino2020virtual}. These include the use of VHs in virtual and augmented reality for e-commerce~\cite{billewar2022rise}, virtual gaming~\cite{verkuyl2017virtual}, visual effects industries and movies~\cite{TurnerWeta2017}, virtual training~\cite{shiradkar2021virtual}, interactions with virtual doctors~\cite{cooks2022can}, sports~\cite{witte2022sports} and virtual try-on for clothing~\cite{zhao2021m3d}.
The need for automatic generation of VHs has therefore been an important research motivation in the last decade which led to the development of several solutions for the creation of high-fidelity VHs. Some methods capture skeletal information from human actors, and use this to animate static geometrically dense 3D VH models (i.e.~meshes or point clouds~\cite{xu2020rignet,chen2021snarf}). Other methods directly capture dense surface data from actors using 4D acquisition platforms based on 4D reconstruction methods~\cite{armando20234dhumanoutfit,wu20244d, mildenhall2021nerf, zins2023multi}. Finally, with the advance of data-driven methods such as generative models~\cite{zhang2022avatargen}, diffusion models~\cite{huang2024humannorm}, and VH retargeting methods~\cite{rekik2024correspondence, jiang2022h4d}, it is now possible to generate new VHs based on geometrically dense data. 

Although the creation of high-fidelity appearance has been developed, the motion of the VH can introduce multiple errors in the geometrically dense VH animation. Thus, assessing the animation quality of such generated VHs remains challenging. One commonly used method consists of measuring the difference between generated and ground truth animations, often captured from human actors, using either objective or subjective measures~\cite{rekik2024survey}. %
For objective evaluations, quantitative metrics are computed, which typically focus on evaluating a specific aspect of the VH including geometrical details~\cite{wang2020neural}, skeletal motion~\cite{zhang2023skinned}, or body parts reconstruction~\cite{yu2023acr}. For subjective evaluations, experiments involving human participants are conducted using perception metrics, such as questionnaires~\cite{voas2023best, wang2024aligning}, physiological and behavioral measures~\cite{herrera2020effect,aburumman2022nonverbal} during participant and VH interactions. 

Recent works~\cite{voas2023best,wang2024aligning} leveraged human ratings to train neural networks to propose evaluation metrics that assess the quality of VH animations. However, these metrics are task-oriented. 
Voas et al.~\cite{voas2023best} suggest a metric that evaluates the faithfulness of a generated motion with respect to a text prompt, while Wang et al.~\cite{wang2024aligning} propose a metric for parameterized motions. The latter metric is trained on data generated with a parametric body model~\cite{bogo2016keep} and can therefore only evaluate the naturalness of parametric human motion. 
To our knowledge, there is currently no objective metric that can provide perceptually meaningful evaluations of non-parametric geometrically dense animated VHs. The reason for this lack is two-fold. First, there is a lack of datasets of geometrically dense animated VHs with subjective evaluation scores. Second, there is no perceptually validated objective metric to globally evaluate the quality of VHs. In this work, we address this problem by generating a dataset of VH animations for which we collect subjective evaluation scores in a user study. We then use the resulting data to propose a first objective quality assessment measure that predicts perceptual evaluation scores in a data-driven framework. Our approach is inspired by works that developed large perceptual datasets and metrics for human faces~\cite{wolski2022geo} and 3D models~\cite{nehme2020visual,nehme2023textured}.

To generate a meaningful dataset of VH animations for subjective annotations in user studies, we distort a reference animation according to several dimensions, inspired by various common artefacts in automatically generated VH animations. Unlike the state of the art methods~\cite{voas2023best,wang2024aligning} that only alter the locomotion of generated VHs, the originality in our study is focusing on both geometry and motion (global and local) of a single non-verbal VH. We evaluate the perceived differences between a \emph{generated VH} and an \emph{acquired VH}. The \emph{acquired VH} is captured using a dense markerless motion capture system. The \emph{generated VH} is derived from the acquired VH by introducing different types of distortions, typically encountered in animation, ranging for slight to strong.

We use the resulting pairs of corresponding \emph{generated VH} and \emph{acquired VH} animations in a perceptual user study to obtain subjective quality scores. For each stimulus, we calculate the mean opinion score (MOS) based on the ratings from all participants. This results in the  \emph{4DHumanPercept} dataset, the first dataset of VH animations acquired using a 4D acquisition system and distorted along controlled factors with corresponding perceptual similarity labels. The dataset is composed of a training and validation dataset (240 stimuli) and a test dataset (10 stimuli). The former involves 240 stimuli created from 8 acquired reference animations with different actors (1 female, 1 male), motions (walk, hop) and clothing (tight, loose), each distorted by 6 error types in 5 distortion levels each. The latter is composed of 10 stimuli resulting of applying randomly one level of distortion on 8 new acquired reference animations coming from 5 subjects (2 female, 3 male) in either tight or loose outfits, exhibiting the motions walk or hop.

We then use the \emph{4DHumanPercept} dataset to understand the factors that impact the quality of the perceived generated VH animations. We further compute a quality measure, called \emph{4DHumanQA}, using a data-driven approach that operates both on the mesh and skeletal domains, to capture geometric and motion distortions, respectively. \emph{4DHumanQA} is a linear combination of geometric and motion-related perceptually significant characteristics of a VH, optimised using subjective scores from \emph{4DHumanPercept}.

The contributions are: 

\begin{itemize}
    \item A dataset of 250 animated VHs with their corresponding MOS, the result of 24 subjects' ratings of each stimulus. This is the first dataset for quality assessment composed of 3D human animations of acquired VHs, distorted with the most common errors in VH generation. The code and dataset are publicly available for research purposes at \url{https://gitlab.inria.fr/rrekikdi/4dhumanqa} .

    \item An analysis of the effects of different acquired reference VHs, as well as distortion types and strengths on MOSs. 
    \item An evaluation of the correlation between a set of perceptually-relevant geometry-based and motion-based features in a 3D human animation with human perception, i.e.~MOS.
    \item The first perceptually-validated quantitative measure for 3D human animation quality assessment. This data-driven method evaluates non-parametric VH animations quantitatively on both geometry and motion levels with human judgement in the loop. 
\end{itemize}

%% file: sections/related-work.tex
\section{Related works}
In this section, we review prior research on the evaluation of VH animations, and discuss relevant evaluation metrics for other graphical content.

\subsection{Evaluation of virtual human animations } 

Inspired by the recent survey by Rekik et al.~\cite{rekik2024survey}, we categorize VH animation evaluation studies into three main types: objective, subjective, and hybrid evaluations.
Depending on the focus of each study, the assessment of VH animation quality may relate to various aspects, such as the realism of global or local motion, geometric detail or physical plausibility. 
\subsubsection{Objective quality assessments} 
\emph{Generated VH} animations can be evaluated quantitatively by comparing them with \emph{acquired VH} animations (e.g., by computing distances between the acquired and the generated VH), or by evaluating whether they respect pre-defined human motion laws.
 
In case of spatially sparse data (i.e.~skeletons), commonly used metrics include Mean Per Joint Position Error (MPJPE), Procrustes aligned MPJPE (PA-MPJPE), which eliminates the error in global displacement~\cite{xu2024finepose,zhou2024simple}, mean acceleration difference (Acc) and its Procrustes aligned (PA-Acc) version~\cite{rempe2021humor,yang2023qpgesture}. 
In case of geometrically dense data (e.g.~meshes) commonly used metrics include mean-per-vertex distance (MPVD) and Procrustes aligned MPVD (PA-MPVD)~\cite{zhou2024simple} for global extrinsic accuracy of the generated surface evaluation, and mean difference in edge length (MDEL) for the evaluation of the preservation of intrinsic geometry. 
Metrics based on pre-defined human motion laws

include the ``two-third power law'' between velocity and curvature~\cite{viviani1995minimum, pham2007formation}, person-ground contact~\cite{rempe2021humor}, Physical Foot Contact (PFC)~\cite{tseng2023edge} or physical plausiblity by using musculoskeletal model simulation resulting from biomechanics research~\cite{geijtenbeek2010evaluating}. 

However, these automatic evaluation metrics cannot effectively reflect or replace subjective user studies, which are crucial to evaluate the \emph{generated VHs} as their primary purpose is to be visually perceived and interacted with by human users.

\subsubsection{Subjective quality assessment} 
Evaluating \emph{generated VHs} with humans in the loop has been done through self-report studies by asking human users to rate \emph{generated VHs} using Likert scales~\cite{justice2022we,rekik2024correspondence}, or through behavioural user studies evaluating user reactions to VHs, notably in immersive environments~\cite{zibrek2019photorealism,subramanyam2020comparing}.

\subsubsection{Hybrid quality assessments} 

Hybrid evaluations quantify the level of realism of VH animations by including human perception in the loop. 

First, works such as~\cite{reitsma2003perceptual, ren2005data, luo2012perceptual, daniel2021perceptually} presented optimised metrics that were based on the correlation between user perception of VH realism and objective metrics to quantify its level. Each work focused on one aspect to evaluate in the VH animation, e.g.~physical balistic~\cite{reitsma2003perceptual} or global trajectory~\cite{daniel2021perceptually} realism.

Recently, with the surge of data-driven methods that train neural networks, novel metrics have been trained on large datasets of subjective ratings using DL-based architectures such as~\cite{voas2023best, wang2024aligning}. Voas et al.~\cite{voas2023best} introduce MoBERT, a novel metric for text-to-motion generation, focusing on naturalness and faithfulness of VHs. They propose a subjective dataset composed of 1400 motion-text pairs with human ratings and use it for the training of their neural network architecture, which is composed of a single multimodal transformer encoder. The input of the proposed data-driven metric is the text prompt and the generated motion, and the output is the perceptual score. 
Wang et al.~\cite{wang2024aligning} present MotionPercept, which is a large-scale human perceptual evaluation dataset containing pairs of human preference annotations on generated motion, and MotionCritic, which is a model trained on the MotionPercept dataset to automatically judge motion quality in alignment with human perceptions. MotionCritic is trained and evaluated on parametric models, more precisely, SMPL motion~\cite{loper2023smpl} represented by 24 axis-angle rotations and one global root translation.
It however does not assess the quality of \emph{generated VHs} in terms of geometric detail preservation with respect to \emph{acquired VHs}.

There are no metrics trained on ratings that evaluate the different aspects of non-parametric \emph{generated VHs} including their local and global motions and geometrical details.

\subsection{Perceptually-validated evaluation methods of other graphical content} 

Although there are few perceptually validated metrics to evaluate animated VHs, there is more research on such metrics for other graphical content, from which we took inspiration for evaluating VH animations.

\subsubsection{Evaluation of static content}
For 2D and 3D graphical content, both traditional and data-driven approaches have been developed, with a growing focus on neural methods that leverage large subjective datasets.
Numerous metrics exist to evaluate the quality of 2D or 3D images, that can be divided into traditional (e.g.~\cite{wang2004image}) and data-driven metrics that include human inputs in the evaluation loop, e.g.~for image~\cite{zhang2018unreasonable,bhardwaj2020unsupervised,ghildyal2022shift,ding2020image,zhu2022deepdc} or 3D mesh~\cite{nehme2019comparison,nehme2023textured} evaluation). 

First, the pioneer work LPIPS~\cite{zhang2018unreasonable} trains a neural network using human-rated similarity dataset of images and uses distances in feature space as perceptual metric. This correlates well with human perception of image similarity. Follow-up works suggest task-oriented metrics such as PIM~\cite{bhardwaj2020unsupervised} for unsupervised evaluation of image similarity, variations of LPIPS~\cite{ghildyal2022shift} to evaluate perceptual similarity in the case of small image misalignments, and DeepDC~\cite{zhu2022deepdc} that suggests a metric without relying on fine-tuning with Mean Opinion Scores (MOS). To evaluate 3D meshes, Nehme et al.~\cite{nehme2019comparison} introduce a dataset consisting of pairs of 3D meshes and corresponding MOS based on human ratings. They use this dataset to analyze how various distortions applied to ground truth meshes affect the ratings, explore the correlation between MOS and objective metrics, and propose an optimized linear model, which is a linear combination of these metrics trained using human ratings. This approach was subsequently extended~\cite{nehme2023textured} by proposing a larger dataset, and training a deep learning-based LPIPS-inspired metric for 3D mesh evaluation that aligns with human perception.

\subsubsection{Evaluation of 4D content}

Perceptual metrics for 4D content, such as video quality assessments, have evolved to incorporate the temporal aspect that distinguishes videos from static data. These metrics aim to evaluate how changes over time impact the perceived quality, considering factors like motion smoothness, temporal consistency or frame rate. The survey of Min et al.~\cite{min2024perceptual} on video quality assessment provides a comprehensive review of both classical and recent approaches in this field. As an example, Hou et al.~\cite{hou2022perceptual} proposed a perceptual quality metric specifically designed to evaluate the quality of interpolated video frames, taking into account both spatial and temporal characteristics. 

%% file: sections/overview.tex
\section{Overview}

The previous section demonstrated a lack of perceptually-validated quantitative quality measures for geometrically dense generated VHs. 

Taking inspiration from previous work on perceptual metrics for static 3D models~\cite{nehme2019comparison,nehme2023textured}, this work introduces a quality measure to predict perceptual assessment scores of geometrically dense animations of VHs.

Our methodology is detailed in Figure~\ref{fig:Teaser}, and organized according to the following steps. Our first objective is to design and release a dataset\footnote{https://gitlab.inria.fr/rrekikdi/4dhumanqa} composed of 3D human animations, or ``4D humans'', with distortions along different axes and levels, with their corresponding user perceptual ratings. This dataset will be referred to as \textit{4DHumanPercept} in the following. The creation of the 3D human animations is detailed in Section~\ref{sec4}, including the different types and degrees of distortions. Section~\ref{sec5} presents the subjective experiment we conducted to acquire the perceptual ratings of the different 3D human animations, and its associated results. %
Our second objective is to design and validate a novel quality measure to predict perceptual evaluation scores of 4D humans, based on this dataset. This novel measure 
called \emph{4DHumanQA} is presented and compared to a state of the art DL-based baseline in Section~\ref{sec7}.

%% file: sections/stimuli-generation.tex
\section{Stimuli generation}\label{sec4}

This section provides details on the generation of the 3D animations used in the subjective evaluation. These animations are based on 4DHumanOutfit~\cite{armando20234dhumanoutfit}, a dataset of densely sampled spatio-temporal 4D human motion data of different actors in different outfits and motions. 

The generated animations are annotated to support the training and testing of data-driven models. For the training and validation dataset, we used a subset of 8  acquired VH models as source characters. Each was distorted using 6 types of distortion, with 5 levels of severity per type, resulting in a total of 240 generated VH animations.
For the test dataset, we used 10 different source models, and each distorted along one dimension. The following sections provide details on the source models, the types of distortions applied, and how these distortions were computed. 

\subsection{3D source model selection}\label{sec:sources}
\begin{figure}
    \centering
    \includegraphics[width=0.5\textwidth]{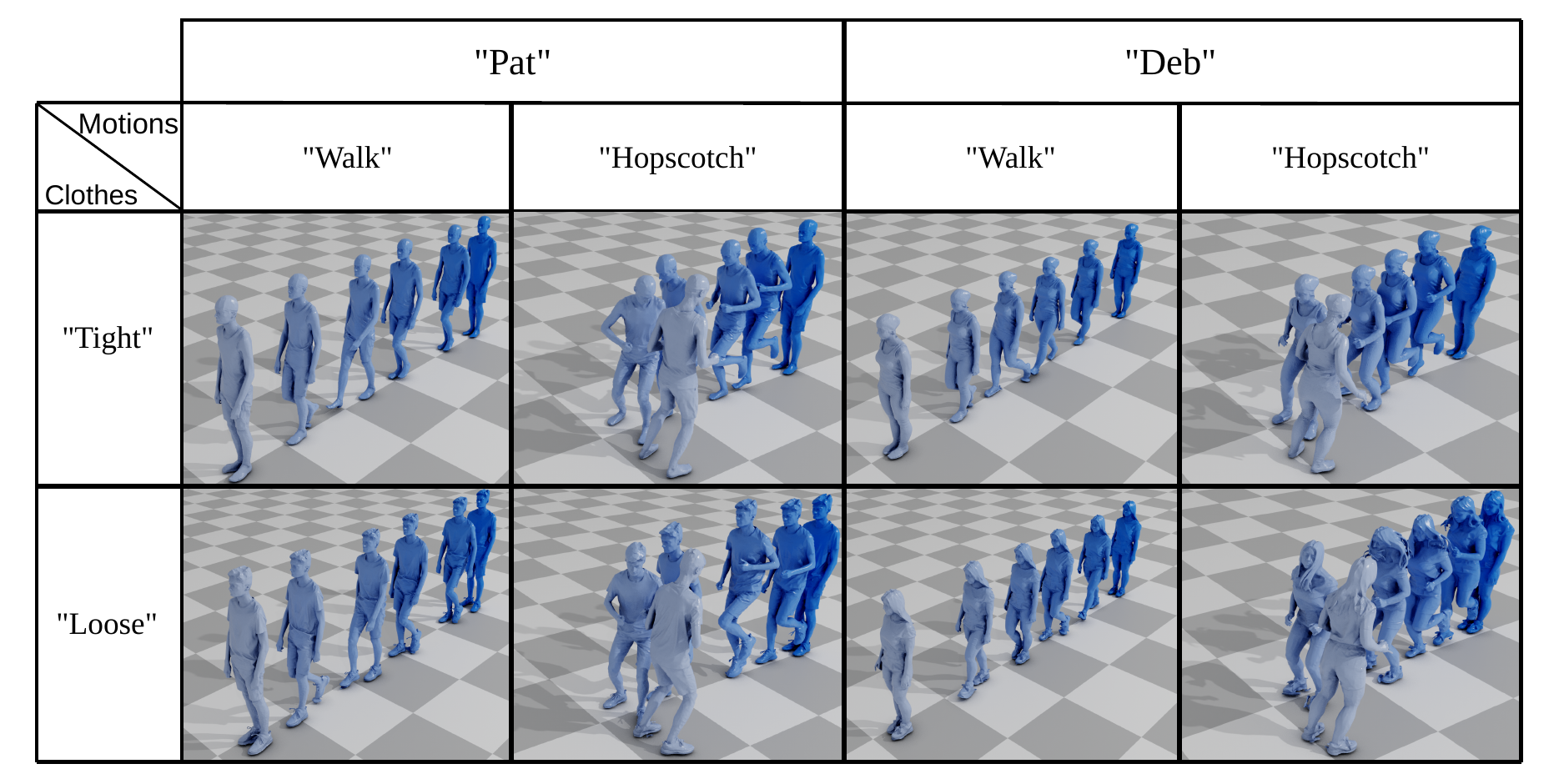}
    \caption{Illustration of the 8 source models selected from 4DHumanOutfit~\cite{armando20234dhumanoutfit}.
    }
    \label{fig:source_models}
\end{figure}

Our goal is to evaluate the quality of generated VHs independently of the method that synthesized them. For the training and validation dataset, we achieve this by distorting 8 chosen source models, corresponding to two subjects (a female with smaller height and heavier build \emph{deb} and a male with taller height and lean build \emph{pat}) in two different outfits each (namely minimal clothing \emph{tig}, and sneakers, shorts and T-shirt \emph{sho}). In our experiments, we consider two motions: \emph{walk}, which is a cyclic motion with large global trajectory changes, and \emph{hopscotch}, which is a non-cyclic motion exhibiting both global trajectory and varied local motions.  
This subset is chosen to contain variety in body shape, clothing, and motion. Figure~\ref{fig:source_models} illustrates the 8 acquired VHs we use as source models.

For the test dataset, we selected 8 source models corresponding to 5 new subjects with varying heights and builds (\emph{ada}, \emph{bea}, \emph{joy}, \emph{tom}, \emph{mat}), each performing either a \emph{walk} or \emph{hop} motion while wearing either a \emph{tig} or \emph{sho} outfit.

Our source models are sequences of 3D human meshes that have neither spatial correspondences between anatomically corresponding body parts, nor temporal correspondences between corresponding frames in similar motions. We use the most detailed version of the reconstructed data to have the best possible geometric and motion details, which are then down-sampled using a mesh simplification algorithm based on the quadratic error metric and triangle collapse~\cite{garland1997surface} to reduce the complexity of the 3D model while preserving its overall shape and appearance as much as possible.

\subsection{Distortions}
\label{sec:distortions}

We consider 6 distortion types, each affecting either global motion, local motion, or geometry. These distortion types are not combined and for each type, we propose 5 strengths.

\subsubsection{Selection of distortion types}

We distort the source models using 6 distortion types applied on geometry, global motion and local motion. The following distortions are chosen according to two criteria. First, to be representative of artifacts which frequently occur and can significantly impact the realism of generated VHs. Second, for their capacity to be simulated fully automatically. 

To alter global motion, we introduce \textbf{footskating}~\cite{Kovar2002,pravzak2011perceptual,zou2020reducing}, a common artifact in motion generation, which can be divided into either foot \textbf{``sliding''}, where the foot slides along the floor while maintaining contact, or \textbf{``moonwalking''}, where the foot slides backwards. We also simulate \textbf{foot contact} problems~\cite{villegas2018neural} (such as foot floating), generally occurring when the character's feet fail to properly interact with the ground plane. \textbf{Motion smoothness} distortions~\cite{mao2021generating} were also added, affecting the overall fluidity of movement, potentially resulting in jerky or unnatural transitions between poses. 

To alter local motion, we add \textbf{twist artefacts}~\cite{rumman2017skin}. They appear around joints, causing unrealistic deformations in areas like the shoulders, hips or legs.

To alter geometry, we include \textbf{self-intersections}~\cite{mihajlovic2022coap,wu2019novel,davydov2024cloaf}, one of the most common errors in VH animation, where different body parts inappropriately overlap or penetrate each other, compromising the physical plausibility of the 3D model. 

\subsubsection{Distortion process}

We use acquired VH source models and distort them. As these models have neither spatial nor temporal correspondences, we proceed by registering all models to a parametric body model, deforming this model to simulate the selected distortion types, and deforming the acquired VHs to be close to the parametric body model's surface. 

\paragraph*{Registration to a parametric model} To register the acquired VH source models, we fit a parametric human body model to each frame. In our implementation, we use SMPL~\cite{loper2023smpl}, which has three sets of parameters to represent the human body. Shape parameter $\beta$ describes an individual's morphology, pose parameter $\theta$ controls the 3D rotations of the kinematic skeleton, and translation parameter $\gamma$ presents the translation of the root of the skeleton.

Let the animation of the acquired VH $\mathcal{A}$ be a sequence of $n$ scans $\{\mathcal{S}_i^A\}_{i=1}^{n}$. We denote the SMPL model fitted to $\{\mathcal{S}_i^A\}_{i=1}^{n}$ by $\{\mathcal{F}_i^A\}_{i=1}^{n}$ and its surface represented by a registered triangle mesh by $\{\mathcal{T}_i^A\}_{i=1}^{n}$. We also denote the SMPL model fitted to $\{\mathcal{S}^A\}_{\mbox{Tpose}}$, which is $\mathcal{S}^A$ in T-pose, by $\{\mathcal{F}^A\}_{\mbox{Tpose}}$ and its corresponding surface by $\{\mathcal{T}^A\}_{\mbox{Tpose}}$.

First, we use $\beta^A$ of $\mathcal{A}$ provided with 4DHumanOutfit. We optimize $\{\mathcal{T}_i^A\}_{\mbox{Tpose}}$ to be as close as possible to $\{\mathcal{S}_i^A\}_{i=1}^{n}$ to predict $\{\mathcal{T}_i^A\}_{i=1}^{n}$ by minimizing a distance loss 
{\small
\begin{equation}
        \mathcal{L}_{\mbox{dist}} =  \lambda_{\mbox{chamfer}}\mathcal{L}_{\mbox{chamfer}} + \lambda_{\mbox{cloth}}\mathcal{L}_{\mbox{cloth}}+ \lambda_{\mbox{prior}}\mathcal{L}_{\mbox{prior}}
        \label{eq:loss_fitting}
\end{equation}
}
with weights $\lambda_{\mbox{chamfer}}=10$, $\lambda_{\mbox{cloth}}=0.01$ and $\lambda_{\mbox{prior}}=1$. 

The chamfer distance $\mathcal{L}_{\mbox{chamfer}}$ is computed unidirectionally from $\{\mathcal{S}_i^A\}_{i=1}^{n}$ to $\{\mathcal{T}_i^A\}_{i=1}^{n}$ as
{\small
\begin{equation}
\mathcal{L}_{\mbox{chamfer}} = \sum_{i=1}^{n} GMoF\left(\mbox{dist}_{\mbox{chamfer}}(\mathcal{S}_i^A, \mathcal{T}_i^A)\right)
\end{equation}
}
where $\mbox{dist}_{\mbox{chamfer}}(\mathcal{S}_i^A, \mathcal{T}_i^A)$ is the Chamfer distance between $\mathcal{S}_i^A$ and $\mathcal{T}_i^A$ at frame $i$ and $GMoF(.)$ is the Geman-McClure function.

The clothing term $\mathcal{L}_{\mbox{cloth}}$ is used to ensure that $\{\mathcal{T}_i^A\}_{i=1}^{n}$ remains entirely within $\mathcal{S}^A$~\cite{yang2016estimation} as
{\small
\begin{equation}
       \mathcal{L}_{\mbox{cloth}} = {\sum_{i=1}^n} GMoF\left(\delta \left( \mathcal{T}_i^A (\beta , \theta) - {\mathcal{NN}}\left(\mathcal{T}^A(\beta, \theta),\mathcal{S}_i^A \right) \right)^2 \right)
\end{equation}
}where ${\mathcal{NN}}(\mathcal{T}^A(\beta, \theta),\mathcal{S}_i^A)$ is the nearest neighbor of $\mathcal{T}^A(\beta, \theta)$ on $\mathcal{S}_i^A$, and $\delta$ is set to one if the nearest neighbors are sufficiently close-by with aligned normals and if $\mathcal{T}^A(\beta, \theta)$ is located outside of $\mathcal{S}_i^A$ and to zero otherwise.

To estimate $\theta^A$ and $\gamma^A$ while fixing $\beta^A$, we use as prior loss $\mathcal{L}_{\mbox{prior}}$ a motion prior~\cite{marsot2022representing}. This model encodes a full motion represented by SMPL parameters into a sequence of latent primitives and decodes it into a sequence of body meshes parameterized by $\beta$, $\theta$, $\gamma$. We include this prior in Equation~\ref{eq:loss_fitting} and optimize for latent primitives that lead to $\{\mathcal{T}_i^A\}_{i=1}^{n}$ that best explain $\{\mathcal{S}_i^A\}_{i=1}^{n}$.

\paragraph*{Simulating selected distortions} %
As $\{\mathcal{T}_i^A\}_{i=1}^{n}$ is structured, distortions can be applied on the level of the SMPL parameters automatically.
The acquired VH is distorted via SMPL parameters $\theta^A$ and $\gamma^A$ fitted to $\mathcal{S}^A$ to generate the distorted SMPL parameters $\theta^G$ and $\gamma^G$, which represent the surface $\mathcal{T}^G$. We drop frame indexes in the notation since the distortion is applied per frame.

For global motion, inspired by~\cite{pravzak2011perceptual}, \textbf{footskating} is simulated by manipulating the root joint translation $\gamma^A$ as 
\begin{equation}
   \gamma^G = \gamma^A * \mathcal{K}
\end{equation}
Strength values $\mathcal{K}$ above 1 produce a sliding effect, and values below 1 result in moonwalking.

\textbf{Foot contact} errors are simulated by adding the value of distortion strength $\mathcal{L}$ applied on the vertical axis $\mathcal{L}=(0,0,\mathcal{L}_z)$ to the root joint $\gamma^A$ to generate the distorted $\gamma^G$ as
\begin{equation}
   \gamma^G = \gamma^A + \mathcal{L}
\end{equation}

\textbf{Motion smoothness} distortion is simulated by deleting arbitrary numbers of frames depending on the strength of the distortion, which represents the percentage $\mathcal{S}$ of frames to be deleted. For instance, a $\mathcal{S}=$ $0.5$ distortion strength means that $50\%$ of the frames are randomly deleted.

For local motion, inspired by~\cite{rumman2017skin}, \textbf{twist artefacts} are introduced by rotating specific joints in $\theta^A$ by angle $\alpha$ in areas prone to twisting, such as the feet, to generate the distorted pose parameter $\theta^G$ as   
\begin{equation}
   \theta^G = \theta^A + \alpha
\end{equation}
The strength of the twisting depends on $\alpha$ and the timing of creating the twist was manually adjusted to occur in the middle of the sequence. 

\textbf{Self intersection} is simulated by rotating joints in $\theta^A$ with an angle $\delta$ until parts of the body intersect unnaturally as
\begin{equation}
   \theta^G = \theta^A + \delta
\end{equation}
The intersection volume depends on $\delta$.

\paragraph*{Deforming acquired VHs} 
We deform $\mathcal{A}$ into a distorted model $\mathcal{G}$ that is close to $\mathcal{T}^G$ using SMPL extended into the volume~\cite{bojanic2024pose}.
The first step is to unpose $\mathcal{A}$, i.e.~predict $\mathcal{S}^A_{\mbox{Tpose}}$, and the second step is to repose $\mathcal{S}^A_{\mbox{Tpose}}$ using $\theta^G$ and $\gamma^G$ to generate $\mathcal{G}$. To do so, we use the correspondence between $\mathcal{S}^A$ and its fitting $\mathcal{F}^A$ along with the underlying SMPL skeleton.
To unpose $\mathcal{A}$, Bojani\`{c}~et al.~\cite{bojanic2024pose} consider $\mathcal{F}^A$ an approximation of $\mathcal{S}^A$ with
{\small
\begin{equation}
        \mathcal{F}^A = \mbox{scale} \cdot [\mathcal{W} (\mathcal{T} + \mathcal{B}_S(\beta^A) + \mathcal{B}_P(\theta^A))+ \gamma^A] + \mathcal{V}_{\mbox{offsets}}
        \label{eq:unposing_smpl}
\end{equation}
}
where $\beta^A$, $\theta^A$ and $\gamma^A$ are predicted SMPL parameters for $\mathcal{A}$, and $\mathcal{T}$ are the SMPL template vertices. $\mathcal{B}_S$ and $\mathcal{B}_P$ are SMPL shape blendshapes and pose offsets, $\mathcal{W}$ is the linear blend skinning (LBS) function, 
$\mathcal{V}_{\mbox{offsets}}$ are the vertex offsets of $\mathcal{F}^A$, and $\mbox{scale}$ is a scalar value that modifies the overall size of the SMPL model, which is 1 in our case. 

First, we unpose $\mathcal{F}^A$ into $\mathcal{F}^A_{\mbox{Tpose}}$.
Second, we unpose the scan, i.e.~we predict $\mathcal{S}$ in T-pose, using Equation~\ref{eq:unposing_smpl}. 

To compute correspondences, we employ a straightforward nearest-neighbor approach, where each point of $\mathcal{S}^A$ is matched to its closest neighbor from $\mathcal{F}^A$. As a result, $\mathcal{S}^A_{\mbox{Tpose}}$ can be written as
\begin{equation}
        \mathcal{S}^A_{\mbox{Tpose}} = \mathcal{W'}^{-1} [((\mathcal{S}^{A} -\mathcal{V'}_{\mbox{offsets}}) / \mbox{scale}) - \gamma^A]
\end{equation}
where $\mathcal{W'}$ represents the same LBS function. Similarly, $\mathcal{V'}_{\mbox{offsets}}$ denotes the same vertex offsets as those for $\mathcal{F}^A$.
We exclude $\mathcal{V'}_{\mbox{offsets}}$ from the equation to maintain a shaped scan in its unposed state. The final unposing equation is
\begin{equation}
       \mathcal{S}^A_{\mbox{Tpose}} = \mathcal{W'}^{-1} [(\mathcal{S}^{A} / \mbox{scale}) - \gamma^A]
\end{equation}

To repose $\mathcal{S}^A_{\mbox{Tpose}}$ using the distorted parameters $\theta^G$ and $\gamma^G$, we use SMPL equation
\begin{equation}
\mbox{scale} \times \left[\mathcal{W} \left(\mathcal{S}^A_{\mbox{Tpose}} + \mathcal{B'}_P(\theta^G)\right) + \gamma^G \right] = \mathcal{S}^G
\end{equation}
where $\mathcal{B'}_P(\theta^G))$ are the SMPL pose offsets remapped to $\mathcal{S}^A$ using correspondences.

\begin{figure*}
    \centering
    \includegraphics[width=0.9\textwidth]{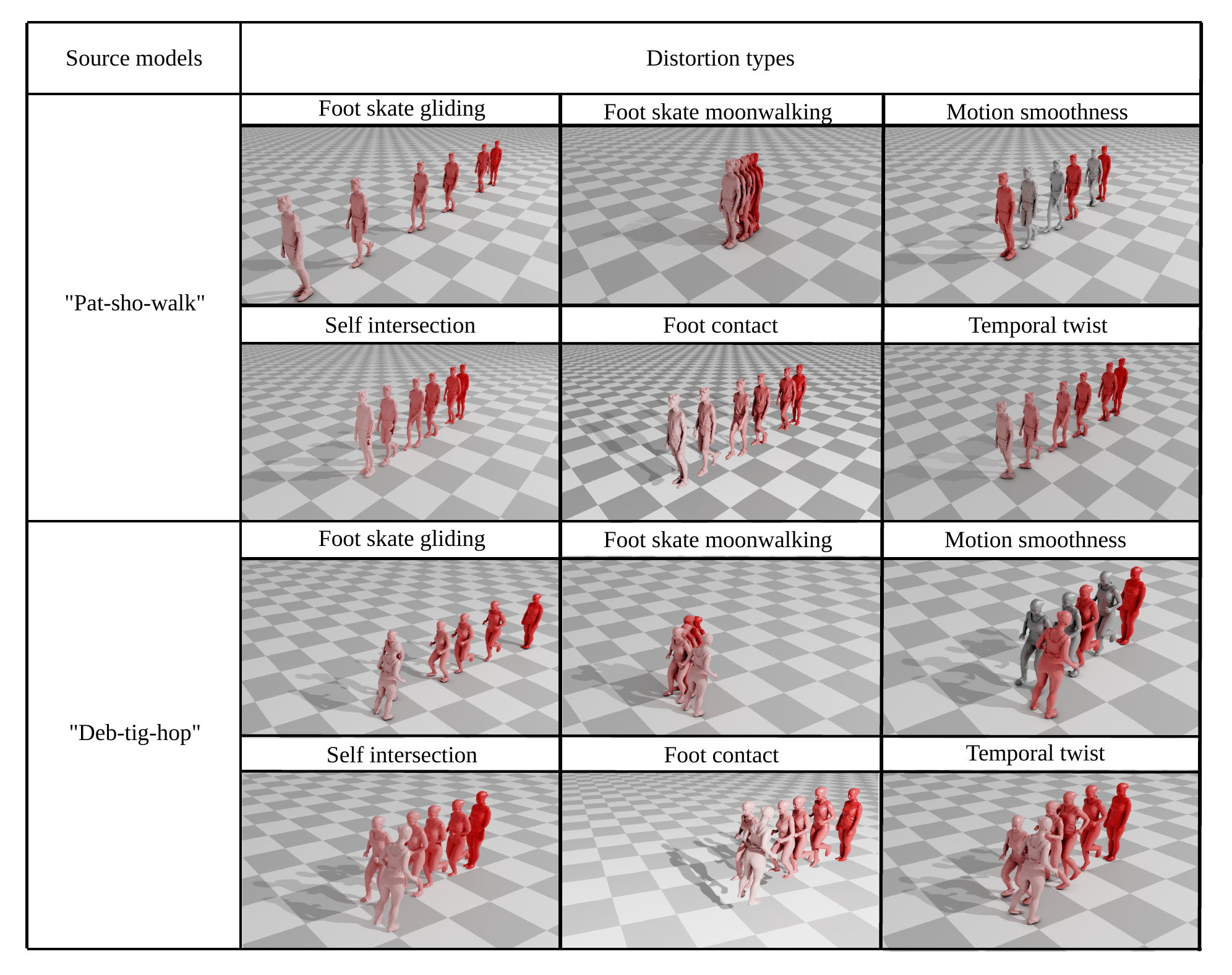}
    \caption{Generated 4D humans with the 6 different simulated distortions at the highest strength. 
    The distortions were applied on two complementary source models, each representing a different subject performing a different motion, and dressed in distinct clothing.
    }
    \label{distortions}
\end{figure*}

To generate the stimuli for the training and validation dataset, the previously detailed deformation process is applied to $\mathcal{A}$ with 6 distortion types, and 5 different strengths each. Figure~\ref{distortions} shows two examples of generated VHs in 2 different outfits exhibiting 2 different motions with the highest level of distortions. 

For the test dataset, each $\mathcal{A}$ was deformed by applying a single distortion type at a single strength level.

%% file: sections/subjective-exp.tex
\section{Subjective experiment}\label{sec5}

The goal of the subjective experiment is to create a dataset of generated VH animations, each labeled with an opinion score that evaluates its realism. The stimuli are the generated and acquired VHs. 
To do so, we use the Double Stimulus Impairment Scale (DSIS) methodology from the International Telecommunication Union (ITU) recommendation~\cite{chalmers2008levels,series2012methodology} as recommended in~\cite{nehme2019comparison}. It consists of showing participants simultaneously the acquired VH (left) and the generated VH (right) stimuli. After watching the videos, participants were asked to answer the following question: ``Please evaluate the visual degradation of the 3D human animation'' using scores ranging from 1 to 5 (where 1: Imperceptible, 2: Perceptible but not annoying, 3: Slightly annoying, 4: Annoying, 5: Very annoying).

The subjective experiment was conducted in two phases. First, a pilot user study was performed to calibrate the distortion levels for each manipulated factor across the different types of distortions. Second, based on the outcomes of the pilot study, we proceeded to the main user study, selecting the appropriate distortion levels. This allowed us to generate 250 perceptually labeled ``generated VHs'' animation samples, with each sample evaluated by 24 participants.

\subsection{Stimuli rendering}
The rendering process for all sequences was done using Blender 4.2. We employed an inclined camera in the top right of the scene to visualize the local movement and global trajectory of the VH. The VH was rendered in a blue color with shadows, enhancing the visibility of foot contact points and grounding the character in the scene. The renders were with width=15.74 inches (1392 pixels) and height=5.89 inches (540 pixels). A checkered background was used to visualize the trajectory and the depth within the scene, offering spatial references. This rendering configuration was maintained across all sequences, ensuring uniformity and facilitating easy comparison between generated and acquired VHs. The experiment was designed using PsychoPy, a convenient and simple Python library for psychological research.
 
\subsection{Design}
\begin{figure}
    \centering
    \includegraphics[width=1\columnwidth]{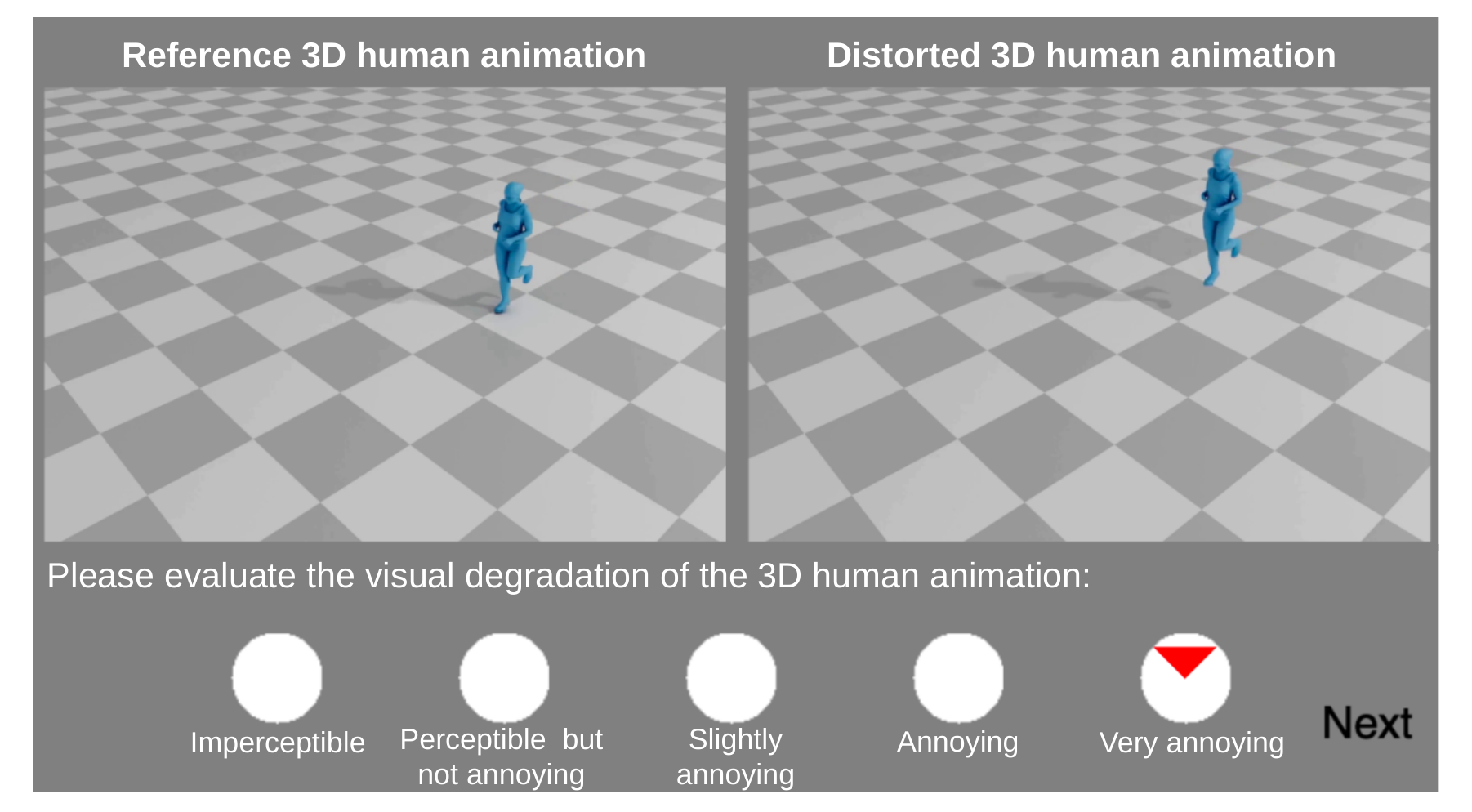}
    \caption{Screenshot of the user study developed with PsychoPy.
    }
    \label{screenshot}
\end{figure}

For both user studies, participants volunteered after being provided with an informative document outlining the details of the experiment. 
They were naive to the purpose of the experiment,
had a normal or corrected-to-normal vision, and gave written and
informed consent prior to the experiment. They were recruited through email lists among students and staff. No compensation was offered. The study conformed to the Declaration of Helsinki, and was approved by the local ethics committee (COERLE). Upon giving their informed consent, participants were seated in front of a 24-inch (16:9) computer screen. They were first asked to complete a questionnaire covering socio-demographic information, including age, gender, and their level of expertise in animation and human motion. Following this, they followed an explanation session and a training session on examples that were not used in the main experiment to get familiar with the task. Afterwards, they proceeded with the experiment, in which they were asked in each trial to observe two videos (the acquired VH one and the generated VH one) and to rate the distorted motion in comparison to the reference motion on a 5-point Likert scale, as illustrated in Figure~\ref{screenshot}. 
 
The stimuli were presented in a randomized order, and participants were not allowed to replay a stimulus once they had submitted their rating. The position of the two videos remained consistent throughout the study: the reference animation was always displayed on the left, and the distorted animation on the right.

\subsection{Pilot user study }
\label{Pilot user study}
 We conducted a pilot study to calibrate the range of each distorsion. The purpose of this pilot experiment was not in getting the exact perceptual thresholds for the individual distortions but to select an appropriate range of animation distortion levels, in terms of minimum and maximum values, as well as appropriate step size. Five participants performed this pilot study: two experts in 4D human data, one expert in 3D modeling and two unfamiliar with VHs.

\subsubsection{Dataset}

The dataset for the pilot study contained a total of 46 stimuli of subject \emph{deb}, in clothing \emph{tig}, exhibiting motion \emph{walk}, to which all 6 distortions were applied. Each distortion was applied with several strength levels, manually chosen to cover a range of non-noticeable to exaggerated distortions with different step sizes. Further details on the strength levels for the distortion types can be found in supplementary material.

\subsubsection{Analysis and results}

Descriptive analysis of the dataset for each individual distortion (interquartile range with medians and max/minimum scores) and non-parametric Friedman analysis with Durbin-Conover pairwise comparisons were conducted to evaluate the perceptual differences between individual step sizes. Further details are provided in supplementary material.

For most distortion types, the step sizes resulted in significantly different estimations in terms of how annoying they were to the evaluators (foot contact: $\tilde\chi^2(6) = 13$, $p = 0.042$; foot skate glide: $\tilde\chi^2(5) = 16.3$, $p = 0.006$; foot skate moonwalking: $\tilde\chi^2(7) = 31.6$, $p < 0.001$; self-intersections: $\tilde\chi^2(5) = 15.7$, $p = 0.008$; temporal twists: $\tilde\chi^2(8) = 29, p < 0.001$). The only exception was motion smoothness, where differences, regardless of the strength, were perceived as less annoying in general (medians mostly 2 and 3, $\tilde\chi^2(8) = 14.7, p = 0.065$). This was also true for foot contact distortion, where the medians for all strengths were around 1 and 2. 

To conclude, based on the statistical analysis, we adjusted our initial set of stimuli to better reflect the appropriate step size, as well as the minimum and maximum values for each distortion type individually. A detailed description of the selection process is provided in supplementary material.

\subsection{Main user study}

\subsubsection{Introduction}

The goal of the main user study is to create a dataset of generated VHs labeled with subjective scores. More specifically, the dataset consists of a main part, called the \textbf{training and validation dataset}, on which we will analyze the Mean Opinion Scores (MOS) and evaluate the influencing factors, and a smaller part, called the \textbf{test dataset}, composed of 10 stimuli, used to test our model in Section VI. Given the number of stimuli in the training and validation dataset (240), making each participant rate all of them would lead to an extremely long experiment duration, which might introduce fatigue and potentially bias our results. Therefore, the type of motion (\emph{walk}, \emph{hop}) is considered as a between factor while ensuring that each participant saw all the distortion types and levels. Therefore, for the training and validation dataset, we chose a mixed design, with a between-subject factor motion type (\emph{walk}, \emph{hop}), and within-subject factors subject identity (\emph{deb}, \emph{pat}), clothing (\emph{tig}, \emph{sho}), distortion type (6 types), distortion strengths (5 levels). Similarly, for the test dataset, we chose a mixed design, with a between-subject factor motion type (\emph{walk}, \emph{hop}), and within-subject factors subject identity (\emph{ada}, \emph{bea}, \emph{joy}, \emph{tom}, \emph{mat}), clothing (\emph{tig}, \emph{sho}), distortion type (1 type per source model), distortion strengths (1 level per distortion).

At the end, each participant was therefore presented with 125 stimuli (120 from the training and validation dataset and 5 from the test dataset).

\subsubsection{Participants} 
Forty-eight participants took part in the experiment.
Participants had different backgrounds with the majority from a research environment. Using a 7-point Likert Scale, participants were asked about their expertise in animation: 16 (33.33\%) were novices (1), 20 (41.67\%) were beginners (2-3), 6 (12.5\%) were intermediate (4), 6 (12.5\%) were advanced (5-6) and none was expert (7). They were also asked about their expertise in human motion: 17 (35.42\%) were novices (1), 16 (33.33\%) were beginners (2-3), 3 (6.25\%) were intermediate (4), 11 (22.92\%) were advanced (5-6), 1 (2.08\%) was expert (7). There were 17 female (35.4\%) and 31 male (64.6\%) participants, aged between 20 and 60 years: 27 (56.25\%) were 20-30 years old, 10 (20.83\%) were 31-40 years old, 5 (10.42\%) were 41-50 years old, and 6 (12.5\%) were 51-60 years old.

\subsubsection{Procedure}
During the training session, participants saw five 8-second stimuli, which were not included in the stimuli to be rated, followed by the rating interface for 5 seconds with the proposed score. The five stimuli were chosen to cover the five strengths of distortions. This was followed by a practice trial stage where 3 extra stimuli were rated by subjects to get familiarized with the task and the rating scale, as suggested in~\cite{series2012methodology}. Results of the training trials weare not used in the subsequent analyses. 
During the core experiment, each participant was then assigned with 125 stimuli to rate (either all walking, or all hopping motions), which corresponds to an experiment duration of approximately 30 minutes.

\subsection{Analysis}
The following analysis thus focuses on the 240 stimuli from the training and validation dataset.
To analyse the scores of a DSIS method, one common way is to compute the Mean Opinion Score (MOS) for each stimulus as  

\begin{equation}
    MOS = \frac{1}{N} \sum_{i=1}^{N} OS_i,
    \label{eq:mos}
\end{equation}
where $N$ is the number of subjects, in our case 24, and $OS_i$ is the opinion score of the $i$-th subject.

To assess the impact of main factors, such as VH identities, VH clothes, VH motions or distortion types and strengths, on the MOS, we conducted 6 separate mixed-design analysis of variance (ANOVA). ANOVA is performed for each distortion type with within-subject factors identity, clothing, and distortion strength, and between-subject factors motion type and participant gender. 
The goal of this comprehensive analysis is to uncover the patterns between the manipulated factors and the participants' opinions, reflected by their scores.
\subsection{Results}\label{sec6}

Table~\ref{tab:ANOVAsGlide} reports the significant results (main and interaction effects) of the ANOVAs for each distortion type. 
We also perform Greenhouse-Geisser correction for violations of sphericity and the effects sizes are reported in the last column~($\eta_p^2$).
Across all distortion types, the distortion strength is the one factor that has consistently a strong effect, which supports the validity of our methodological approach.

\begin{figure*}
    \centering
    \includegraphics[width=1.0\textwidth]{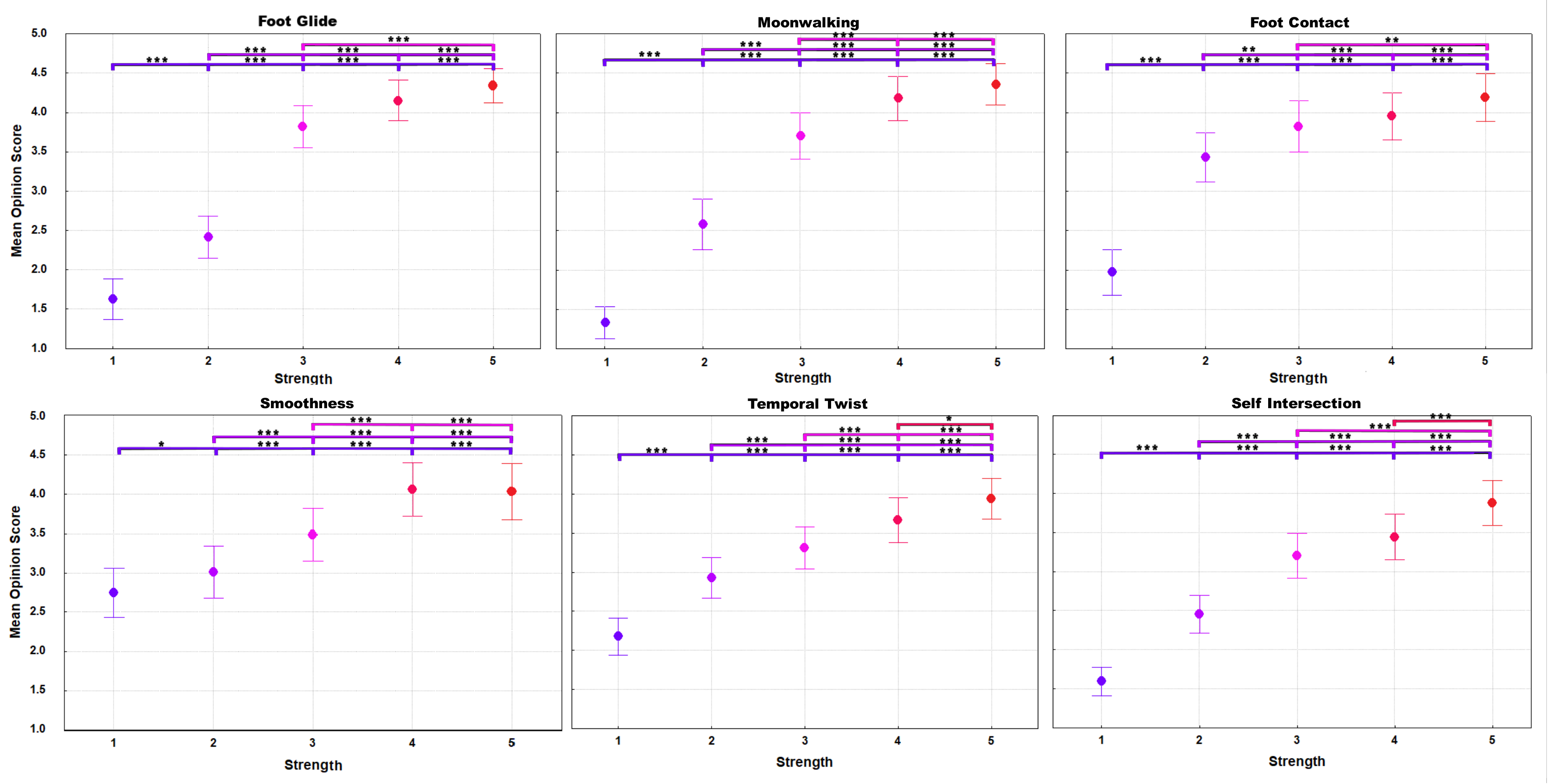}
    \caption{
    Graphs representing the distribution of participants' ``Opinion'' responses with the calculated MOS and 95\% confidence intervals (whiskers) for the factor Distortion strength, for all Distortion types. Lines marked with * denote significant differences at $p < 0.05$, ** $p < 0.01$, and *** $p < 0.001$ (post hoc test: Tukey HSD).
     \label{fig:MOS}
    }
    \label{fig:smpl based deformations }
\end{figure*}

This effect is further illustrated in Figure~\ref{fig:MOS} which shows that the MOS increases with increasing distortion strength for all distortion types. Regarding all the other factors, as shown in Table~\ref{tab:ANOVAsGlide}, we observe that their main effects and interactions vary depending on the type of distortion considered. This variability highlights the complexity of assessing an individual factor's influence on user responses and underscores the importance of a multifactorial approach, where subject identity, motion types, participant gender, and clothing introduce nuances in the perception of VH animation quality.

\begin{table}[!h]
    \centering
    {\footnotesize
    \begin{tabular}{|l|r|r|r|}
    \hline
     \textbf{Effect} & \textbf{F-Test} & \textbf{p-value} & \textbf{$\eta_p^2$}
     \\
    \hline
    \multicolumn{4}{c}{}\\
    \multicolumn{4}{c}{Distortion type \textbf{Foot Glide}}\\
    \hline
    Identity & \Fval{1}{43}{21.5} & 0.000 & 0.33\\
    \hline
    Identity $\times$ Motion Type & \Fval{1}{43}{24.4} & 0.000 & 0.36\\
    \hline
    Identity $\times$ Gender & \Fval{1}{43}{5.7} & 0.022 & 0.12\\
    \hline
    Strength & \Fval[*]{2.47}{106.6}{165.3} & 0.000 & 0.79 \\
    \hline
    \multicolumn{4}{c}{}\\
    \multicolumn{4}{c}{Distortion type \textbf{Moonwalking}}\\
    \hline
    Identity $\times$ Gender  & \Fval{1}{43}{5.00} & 0.031 & 0.10\\
    \hline
    Strength & \Fval[*]{3.0}{129.1}{185.6} & 0.000 & 0.81 \\
    \hline
    \multicolumn{4}{c}{}\\
    \multicolumn{4}{c}{Distortion type \textbf{Foot contact}}\\
    \hline
    Identity $\times$ Gender & \Fval{1}{43}{5.7} & 0.022 & 0.12\\
    \hline
    Strength & \Fval[*]{2.47}{106.6}{165.3} & 0.000 & 0.79 \\
    \hline
    Identity $\times$ Clothing & \Fval{4}{172}{4.4}& 0.002 & 0.09 \\
    $\times$ Strength $\times$ Gender & & & \\
    \hline
    \multicolumn{4}{c}{}\\
    \multicolumn{4}{c}{Distortion type \textbf{Motion smoothness}}\\
    \hline
    Motion Type & \Fval{1}{43}{17.4} & 0.000 & 0.29\\
    \hline
    Gender & \Fval{1}{43}{6.2}& 0.017 & 0.13\\
    \hline
    Identity & \Fval{1}{43}{4.29} & 0.04 & 0.09 \\
    \hline
    Strength & \Fval[*]{18}{78.9}{45.99} & 0.000 & 0.52\\
    \hline
    Strength $\times$ Gender & \Fval{4}{172}{2.96} & 0.021 & 0.06\\
    \hline
    Identity $\times$ Clothing & \Fval{1}{43}{6.27}& 0.016 & 0.13\\
    \hline
    Identity $\times$ Clothing & \Fval{4}{172}{17.1} & 0.048 & 0.05\\
    $\times$ Strength & & & \\
    \hline
    \multicolumn{4}{c}{}\\
    \multicolumn{4}{c}{Distortion type \textbf{Temporal Twist}}\\
    \hline
    Motion Type & \Fval{1}{43}{7.32} & 0.010 & 0.145\\
    \hline
    Gender & \Fval{1}{43}{5.64} & 0.020 & 0.116\\
    \hline
    Identity & \Fval{1}{43}{5.55} & 0.023 & 0.114 \\
    \hline
    Identity $\times$ Motion Type & \Fval{1}{43}{21.27} & 0.000 & 0.33\\
    \hline
    Strength & \Fval[*]{3.2}{138.7}{93.03} & 0.000 & 0.68\\
    \hline
    Identity $\times$ Clothing & \Fval{1}{43}{4.64} & 0.037 & 0.10\\
    $\times$ Motion Type & & & \\
    \hline
    Identity $\times$ Clothing & \Fval{4}{172}{2.99} & 0.001 & 0.06\\
    $\times$ Strength & & & \\
    $\times$ Motion Type & & & \\
    \hline
    \multicolumn{4}{c}{}\\
    \multicolumn{4}{c}{Distortion type \textbf{Self intersection}}\\
    \hline
    Motion Type & \Fval{1}{43}{22.4} & 0.000 & 0.34\\
    \hline
    Gender & \Fval{1}{43}{9.8} & 0.003 & 0.19\\
    \hline
    Identity & \Fval{1}{43}{31.5} & 0.000 & 0.43 \\
    \hline
    Identity $\times$ Motion Type & \Fval{1}{43}{50.8} & 0.000 & 0.54\\
    \hline
    Identity $\times$ Gender & \Fval{1}{43}{6.3} & 0.016 & 0.13\\
    \hline
    Identity $\times$ Gender & \Fval{1}{43}{5.4} & 0.025 & 0.11\\
    $\times$ Motion Type & & & \\
    \hline
    Strength & \Fval[*]{2.5}{108.6}{105.09} & 0.000 & 0.71\\
    \hline
    Strength $\times$ Motion Type& \Fval{4}{172}{9.00} & 0.000 & 0.17\\
    \hline
    Identity $\times$ Clothing & \Fval{1}{43}{7.7} & 0.008 & 0.15\\
    \hline
    Identity $\times$ Clothing & \Fval{1}{43}{17.1} & 0.000 & 0.28\\
    $\times$ Motion Type & & & \\
    \hline
    Clothing $\times$ Strength & \Fval{4}{172}{3.3} & 0.013 & 0.07\\
    \hline
    Clothing $\times$ Strength & \Fval{4}{172}{3.1} & 0.018 & 0.07\\
    $\times$ Motion Type & & & \\
    \hline
    \end{tabular}
    }
    
    \caption{
    Significant main and interaction effects of the independent factors on the variable ``Opinion'', per distortion type. F*~stands for Greenhouse-Geisser correction for violations of sphericity, and effects sizes are reported in the last column ($\eta_p^2$).
    }
    \label{tab:ANOVAsGlide}
   
\end{table}

%% file: sections/metric.tex
\section{Perceptually-validated quality measure for 3D human animation assessment}
\label{sec7}

We aim to predict a perceptual score that can assess the realism level of a generated VH compared to its acquired version. 
First, we introduce perceptually relevant features that could impact the perception of shape and motion of a VH.
Second, we analyse the correlation of each individual feature with MOSs. 
Based on these results, a logistic regression model is trained to predict a perceptual score from a set of perceptually relevant features of a generated VH. The model's parameters were optimized through cross-validation on the subjective data collected during the perceptual experiment.

\subsection{Features for 3D human motion similarity}\label{features}%

The dataset is composed of generated VHs and acquired VHs, which are sequences of 3D unstructured meshes, along with their corresponding MOSs. Each generated or acquired VH is additionally approximated by body shape parameters and skeletal joint positions.
Inspired by a recent survey~\cite{rekik2024survey}, we divide the realism features into geometric and kinematics features, which include global trajectory and local motion.

\subsubsection{Geometric features}

We evaluate shape dissimilarity by computing commonly used measures applied to the full shape, such as the Chamfer Distance and the Hausdorff Distance, which typically assess similarity between point clouds.

 \textbf{Chamfer distance} (feature $F_1$)~\cite{fernandes2021point}  is defined as
    {\footnotesize
\begin{eqnarray}
    F_1(\{\mathcal{S}_i^A\}_{i=1}^{n}, \{\mathcal{S}_i^G\}_{i=1}^{n}) & = & \frac{1}{n} \sum_{i=1}^{n} \left( \frac{1}{|\mathcal{S}_i^A|} \sum_{x \in \mathcal{S}_i^A} \min_{y \in \mathcal{S}_i^G} \|x - y\|^2 \right. \nonumber \\
    & & \quad + \left. \frac{1}{|\mathcal{S}_i^G|} \sum_{y \in \mathcal{S}_i^G} \min_{x \in \mathcal{S}_i^A} \|x - y\|^2 \right)
\end{eqnarray}

}

 \textbf{Hausdorff distance} (feature $F_2$)~\cite{van2022between} 
is defined as

{\footnotesize
\begin{eqnarray}
    F_2(\{\mathcal{S}_i^A\}_{i=1}^{n}, \{\mathcal{S}_i^G\}_{i=1}^{n}) & = & \frac{1}{n} \sum_{i=1}^n
    \max \Bigg\{ \max_{x \in \mathcal{S}_i^A} \min_{y \in \mathcal{S}_i^G} \|x - y\|, \nonumber\\
    & & \max_{y \in \mathcal{S}_i^G} \min_{x \in \mathcal{S}_i^A} \|x - y\| \Bigg\}
\end{eqnarray}
}

\subsubsection{Kinematic features}

Kinematic features include measures on global trajectory and on local motion.

\textit{Global trajectory features} include four different types of features.

    \textbf{Foot contact} (feature $F_3$), where we compute the difference of the feet stability between corresponding frames of $\mathcal{A}$ and $\mathcal{G}$, and average through the sequence as
\begin{equation}
F_3 = \frac{1}{n} \sum_{i=1}^n 
\left( \|\gamma_{i}^G - \gamma_{i}^A\|  \right),
\end{equation}
where $\gamma_{i}^A$ and $\gamma_{i}^G$ are the translations of root joints at frame $i$ in $\mathcal{A}$ and $\mathcal{G}$, respectively. 

\textbf{Global translation} (feature $F_4$) is computed as the average displacement between the root joints of $\mathcal{A}$ and $\mathcal{G}$~\cite{shin2024wham} as
\begin{equation}
F_4 = \frac{1}{n} \sum_{i=1}^n \left( \|p_{i}^G - p_{i}^A\|  \right),
\end{equation}
where $p_{i}^A$ and $p_{i}^G$ are the positions of root joints at frame $i$ in $\mathcal{A}$ and $\mathcal{G}$.

 \textbf{Difference in motion velocity} (feature $F_5$) between $\mathcal{A}$ and $\mathcal{G}$, computed as 
{\footnotesize 
\begin{equation}
   F_5 = \left| \frac{1}{n-1} \sum_{i=1}^{n-1} \frac{\mbox{dist}(\mathcal{S}_i^A, \mathcal{S}_{i+1}^A)}{\mathcal{D}_A}
    - \frac{1}{n-1} \sum_{i=1}^{n-1} \frac{\mbox{dist}(\mathcal{S}_i^G, \mathcal{S}_{i+1}^G)}{\mathcal{D}_G} \right|,
\end{equation}}
where $\mbox{dist}$ is the nearest neighbour distance between two successive frames in $\mathcal{A}$ or $\mathcal{G}$, and $\mathcal{D}_G$ and $\mathcal{D}_A$ are the durations of sequences $\mathcal{G}$ and $\mathcal{A}$, respectively.

\textbf{Motion smoothness} (feature $F_6$) is inspired by~\cite{gulde2018smoothness,bayle2023comparison} and defined as
\begin{equation}
F_6 = \frac{1}{j} \sum_{i=1}^{j} \left( \mbox{LDLJ}^A_i - \mbox{LDLJ}^G_i \right),
\end{equation}
where $j$ is the total number of SMPL joints, and $\mbox{LDLJ}$ is the log dimension-less jerk, defined as
\begin{equation}
\mbox{LDLJ} = -\ln\left( \frac{(t_2-t_1)^5}{L^2} \int_{t_1}^{t_2} \left( \frac{d^3x}{dt^3} \right)^2 dt \right)
\end{equation}
where $t_1$ and $t_2$ are the start and end time of the sequence, $L$ is the path length, i.e.~the total distance traveled along the trajectory, $x(t)$ is the position vector, and $\frac{d^3x}{dt^3}$ is the jerk, which is the third derivative of position with respect to time.

\textit{Local motion features} are features that consider individual joint positions. We use one feature belonging to this category, denoted by $F_7$, which is the \textbf{mean per joint position error} (MPJPE) between $\mathcal{A}$ and $\mathcal{G}$ defined as
\begin{equation}
    F_7 = \frac{1}{j \cdot n} \sum_{t=1}^n \sum_{i=1}^j \| p_{i,t}^{A} - p_{i,t}^{G} \|_2,
\end{equation}
where $p_{i,t}^{A}$ and $p_{i,t}^{G}$ are positions of the $i$-th joint at frame $t$ of $\mathcal{A}$ and $\mathcal{G}$, respectively.

\subsection{Single feature prediction performance}\label{single features}

To predict how realistic viewers would perceive animations to be, we analyzed the correlation across the entire dataset between the scores of realism features and the ground truth MOS obtained from the user study. For each video, we extracted the average value of each metric and evaluated the correlation with the corresponding MOS scores. Since the data distribution was not normal (as confirmed by the Shapiro-Wilk test, $p < 0.05$), we computed the Spearman Rank Order Correlation Coefficient (SROCC) 
along with its associated $p$-value.

Table~\ref{tab:features_table} shows the results of the Spearman rank order correlation analysis separately for each feature. While all correlations are significant, their strength is small to medium, highlighting the need to consider more complex metrics which combine several features.

\begin{table}
    \centering
    {
    \small
    \begin{tabular}{|c|c|c|c|}
    \hline
    Feature & ID & SROCC & $p$-value \\ 
    \hline
    \hline
    Chamfer distance & F1 & 0.156 & 0.016* \\ 
    \hline

    Hausdorff distance & F2 & 0.177 & 0.006*\\
    \hline
   
    Foot Contact & F3 & 0.144  & 0.025*\\ 
    \hline
   
    Global translation  & F4 & 0.389 &  $<$0.001*\\ 
    \hline
  
    Velocity & F5 & 0.255  & $<$0.001* \\
     \hline
    Motion smoothness & F6 &  0.267 & $<$0.001* \\ 
    \hline
    Mean per-joint position error & F7 & 0.436 & $<$0.001* \\ 
    \hline
   
    \end{tabular}
    }

    \caption{Spearman correlation analysis between features and MOS. Significant correlations are highlighted with a *.}
    
    \label{tab:features_table}
\end{table}

\subsection{Quality assessment of 3D human animation}
\label{model}

Our goal is to develop a quality assessment measure that better correlates with human perception. To this end, we propose a new model based on perceptual features and evaluate its ability to predict observer scores.

The proposed model is evaluated on test data and compared to a state of the art deep learning baseline. 
Our \emph{4DHumanPercept} dataset is composed of 250 pairs 

of acquired VHs and generated VHs along with corresponding MOSs. For each data point, we have 7 objective features. 
We develop a data-driven model, trained and tested on our dataset, to find the best combination of features for MOS prediction. 

\subsubsection{Perceptually-optimised metric}\label{metricmodel}

The proposed quality assessment measure, called \emph{4DHumanQA} is a linear regression model trained to predict the MOS. The input of the model are the 7 features describing the objective distances between the acquired and generated VHs and the output is the predicted MOS by minimising mean squared error (MSE). The model is trained using the training (80\% of 240) and validation (20\% of 240) dataset and tested using the test dataset (10 videos).

\subsubsection{Results}

For each pair of VHs from the test dataset, we computed features $F_i, i=1,\ldots,7$, and used the linear regression model to predict the MOS. This value can be compared to the MOS values collected during the user study. We compare the predicted and collected MOS in terms of MSE, Pearson Linear Correlation Coefficient (PLCC), and SROCC.

\begin{table}[h]
    
    \centering
    {\small
    \begin{tabular}{|c|c|c|}
        \hline
        Metrics & 4DHumanQA (Ours) & LPIPS~\cite{zhang2018unreasonable} \\
        \hline
        \hline
        Mean squared error & \textbf{0.178} & 0.515\\

        \hline
        PLCC & \textbf{0.917} &  0.729 \\
        ($p$-value) & (1.89e-4) & (1.70e-2)\\
        \hline
        SROCC & \textbf{0.961} & 0.76 \\
        ($p$-value) &  (1.00e-5) & (1.10e-2) \\
        \hline
    \end{tabular}
    }
    \caption{Comparison of MOS collected during the user study and predictions from regression model \emph{4DHumanQA} and LPIPS pretrained model~\cite{zhang2018unreasonable} on test data. Best scores are in \textbf{bold}.}
    \label{tab:my_label}
\end{table}

Table~\ref{tab:my_label} shows the results. Predicted MOSs by our model present strong correlations, SROCC and PLCC, with collected MOS ($>0.9$) and both correlations are significant. The MSE is low which implies the model is accurate. These results demonstrate that \emph{4DHumanQA} predicts MOS of unseen animated VHs well.
 
\paragraph*{Comparison to LPIPS }
\label{sec8}
We compare \emph{4DHumanQA} to the deep learning baseline LPIPS~\cite{zhang2018unreasonable}. 
As LPIPS is an image-base metric, we compute this metric on rendered videos of acquired VH and generated VH of the test dataset from \emph{4DHumanPercept} on a per-frame basis, and compute an average predicted MOS scores over all the video frames.

Table~\ref{tab:my_label} shows that \emph{4DHumanQA} outperfoms LPIPS across all evaluation metrics. More precisely, LPIPS presents lower correlation with collected MOSs compared to \emph{4DHumanQA}.

%% file: sections/conclusion.tex
\section{Conclusion }
\label{conclusion}

This work introduced the \emph{4DHumanPercept} dataset, which is the first dataset that contains generated VHs with detailed geometry including hair and layered clothes annotated with subjective scores, describing the visual distortion compared to the corresponding acquired VHs. In addition to detailing the generation process of the stimuli and the subjective experiment, we presented a detailed analysis of the effects of the different source models (subject, clothing, motion), the simulated distortions to create the generated VHs and the strength levels of distortions. Furthermore, we show experimentally that individual quantitative features that present motion and geometry distortion between generated and acquired VHs do not correlate well with human perception. However, the linear combination of all the features, by training a linear regressor supervised with MOS ratings provided in a user study, results in a correlation of $90\%$, which is better than the deep learning baseline LPIPS~\cite{zhang2018unreasonable}. To conclude, this measure can be employed to provide an accurate perceptual evaluation of any geometrically dense 3D human animation.

However, our approach is limited by the choice of stimuli. We only distort two different subjects in two types of clothing exhibiting two motions, which impacts the variability and the size of the subjective dataset and subsequently the proposed data-driven method. Future work includes several axes. First, scaling up the dataset by adding more subjects, clothing, and motions would enable us to increase variability. This can be done by including more sequences of 4DHumanOutfit or other 4D datasets. Second, including more distortions while generating VHs such as shoulder or arm twisting would also result in a more general model. Acquired VHs could also be distorted along the clothing axis by deforming the clothing in an unrealistic way while the VH is moving. This can be challenging, especially if the input consists of unregistered meshes. Inspired by recent work on 3D meshes~\cite{nehme2023textured} and parametric human body models~\cite{voas2023best,wang2024aligning}, collecting a larger volume of subjective ratings—whether by increasing the number of source models or introducing new types of distortions—would greatly benefit the training and validation of deep learning-based metrics that align with human perception. Third, an interesting avenue for future work is to subjectively evaluate the stimuli not only through traditional self-report measures, but also by immersing participants in a virtual reality (VR) environment where they can move freely around the animated VHs. In such a setting, subjective behavioral metrics—such as interpersonal proximity—could provide valuable insights~\cite{rekik2024survey}. For instance, Patotskaya et al.~\cite{patotskaya2023avoiding} introduced a proximity-based measure derived from trajectory analysis, which was shown to be influenced by the agent's animation style. Their findings suggest that virtual agents displaying more unpredictable motion, indirect gestures, and excessive general movement tend to increase users’ discomfort, which manifests as greater personal distance in VR. Building on this, it would be interesting to explore whether animation errors similarly lead to increased proximity distances, thereby offering a novel behavioral indicator of perceived animation quality.
Finally, another promising direction is to replace traditional geometric and kinematic features with deep features learned directly from data. Deep feature extractors such as~\cite{wen2022point, liu2019meteornet, fan2022pstnet, fan2021point} can capture more abstract and context-rich representations of motion, which may better align with human perception. These methods have shown strong performance in processing 3D point cloud data and learning spatio-temporal patterns. Incorporating such deep representations could enhance the accuracy and generalizability of perceptual quality prediction frameworks for virtual human animation.

%% file: sections/Acknowledgments.tex
\ifCLASSOPTIONcompsoc

  \section*{Acknowledgments}
\else
  \section*{Acknowledgment}
\fi

We thank Briac Toussaint for the 3D reconstructions, and Antoine Dumoulin and David Bojani\'{c} for help with the SMPL fittings. This work was partially funded by the French National Research Agency (ANR) 3DMOVE - 19-CE23-0013.

%% file: supplementary/spl.tex
\clearpage
\setcounter{page}{1}
\renewcommand{\thepage}{S\arabic{page}}

\begin{center}
    {\LARGE\bfseries Supplementary Material \par}
  
\end{center}

This supplementary material contains additional explanations of the pilot user study and further details about the linear regression model. 

\subsection{Pilot user study }
\subsubsection{Dataset}

The total number of stimuli rated in the pilot user study is 46. We created the following strength levels for the distortion types (in increasing strength order): 

\begin{itemize}
    
    \item Foot skate gliding: $\mathcal{K}$= 1.115, 1,18, 1.25, 1.5, 1.75, 2
    \item Foot skate moonwalking: $\mathcal{K}$= 0.97, 0.8, 0.75, 0.5, 0.4, 0.3, 0.2, 0.1
    \item Foot contact (m): $\mathcal{L}_z$= 0.025, 0.05, 0.075, 0.1, 0.125, 0.15, 0.175, 0.2 
    \item Motion smoothness: $\mathcal{S}$= 0.1, 0.2, 0.3, 0.4, 0.5, 0.15, 0.25, 0.35, 0.45
    \item Temporal twist: 
    $\alpha$= 0.1, 0.15, 0.2, 0.25, 0.3, 0.35, 0.4, 0.45, 0.5.
    \item Self intersection: $\delta$= 0.05, 0.1, 0.15, 0.2, 0.25, 0.3
\end{itemize}

\subsubsection{Analysis and results}

\begin{figure*}[b]
    \centering
    \includegraphics[width=1.0\textwidth]{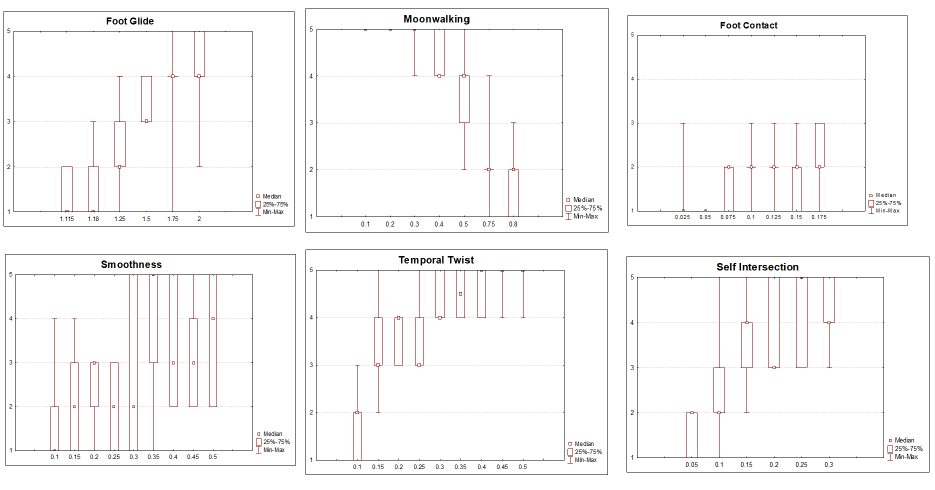}
    \caption{
    Graphs representing the minimum, maximum, and distribution of the evaluators’ answers for each distortion type.
     \label{fig:boxplot}
    }
    \label{fig:boxplot }
\end{figure*}

To select the most suitable distortion strengths for the main experiment, we analysed the minimum, maximum, and distribution of the evaluators’ answers for each distortion type, shown in Figure~\ref{fig:boxplot}. We expected that the appropriate median of answers on the lower bound of the strengths should be 1 or 2 and for the higher strength at least 4 or 5. The distortion levels in between should range from $2$ to $4$. Then, we checked for a stagnation, where the perception of distortion is stable with increasing strength of distortion and pairwise comparisons contain insignificant differences between stimuli strengths.

Most animation distortion types had a distribution of medians starting from 1 or 2 and maximising at 4 or 5. However, for \textbf{self-intersection}, the stagnation of the ratings began at 0.2. To include more intermediate steps and maintain a stable number of strength levels, we adjusted the maximum strength to 0.225 and set the step size to 0.05. For \textbf{motion smoothness}, the perceptual score was not sensitive to small changes of distortion increase. Therefore, intermediate strengths were removed, and the step size set to 0.1 with a maximum of 0.5. For \textbf{temporal twist}, the last four strengths had a median of 5 and the perceived annoyance between them was not significantly different, so they were removed, and the strength stopped at 0.3. \textbf{Foot contact} did not show any spread in answers. We consider larger steps of 0.1 and a maximum of 0.45. The initial strengths for \textbf{foot skate gliding} and \textbf{foot skate moonwalking} were kept, the first two strengths (0.1, 0.2) are perceived almost the same, so we set the maximum to 0.2 and the minimum 0.97, based on~\cite{pravzak2011perceptual}, and the step size to 0.2.

\subsection{Linear regression model}

After training the linear regressor in part \ref{metricmodel}, we obtain the model
\begin{equation}
    M= \sum_{i=1}^7 w_i F_i,
    \label{model}
\end{equation}
with weights $w_1 = 0.246, w_2 = -0.259, w_3 = 0.459, w_4= -7.2, w_5 = -0.273, w_6 = 0.643$, and $w_7 = 7.49$.

%% file: main.bbl
\begin{thebibliography}{10}

\bibitem{aburumman2022nonverbal}
N.~Aburumman, M.~Gillies, J.~A. Ward, and A.~F. d.~C. Hamilton.
\newblock Nonverbal communication in virtual reality: Nodding as a social
  signal in virtual interactions.
\newblock {\em International Journal of Human-Computer Studies}, 164:102819,
  2022.

\bibitem{armando20234dhumanoutfit}
M.~Armando, L.~Boissieux, E.~Boyer, J.-S. Franco, M.~Humenberger, C.~Legras,
  V.~Leroy, M.~Marsot, J.~Pansiot, S.~Pujades, et~al.
\newblock {4DHumanOutfit}: a multi-subject 4d dataset of human motion sequences
  in varying outfits exhibiting large displacements.
\newblock {\em Computer Vision and Image Understanding}, 237:103836, 2023.

\bibitem{bayle2023comparison}
N.~Bayle, M.~Lempereur, E.~Hutin, D.~Motavasseli, O.~Remy-Neris, J.-M. Gracies,
  and G.~Cornec.
\newblock Comparison of various smoothness metrics for upper limb movements in
  middle-aged healthy subjects.
\newblock {\em Sensors}, 23(3):1158, 2023.

\bibitem{bhardwaj2020unsupervised}
S.~Bhardwaj, I.~Fischer, J.~Ball{\'e}, and T.~Chinen.
\newblock An unsupervised information-theoretic perceptual quality metric.
\newblock {\em Advances in Neural Information Processing Systems}, 33:13--24,
  2020.

\bibitem{billewar2022rise}
S.~R. Billewar, K.~Jadhav, V.~Sriram, D.~A. Arun, S.~Mohd~Abdul, K.~Gulati, and
  D.~N. K.~K. Bhasin.
\newblock The rise of 3d e-commerce: the online shopping gets real with virtual
  reality and augmented reality during covid-19.
\newblock {\em World Journal of Engineering}, 19(2):244--253, 2022.

\bibitem{bogo2016keep}
F.~Bogo, A.~Kanazawa, C.~Lassner, P.~Gehler, J.~Romero, and M.~J. Black.
\newblock Keep it {SMPL}: Automatic estimation of 3d human pose and shape from
  a single image.
\newblock In {\em European Conference on Computer Vision}, pages 561--578,
  2016.

\bibitem{bojanic2024pose}
D.~Bojani{\'c}, S.~Wuhrer, T.~Petkovi{\'c}, and T.~Pribani{\'c}.
\newblock Pose-independent 3d anthropometry from sparse data.
\newblock In {\em ECCV Workshop T-CAP}, 2024.

\bibitem{carrozzino2020virtual}
M.~A. Carrozzino, R.~Galdieri, O.~M. Machidon, and M.~Bergamasco.
\newblock Do virtual humans dream of digital sheep?
\newblock {\em Computer Graphics and Applications}, 40(4):71--83, 2020.

\bibitem{chalmers2008levels}
A.~Chalmers and A.~Ferko.
\newblock Levels of realism: From virtual reality to real virtuality.
\newblock In {\em Spring Conference on Computer Graphics}, pages 19--25, 2008.

\bibitem{chen2021snarf}
X.~Chen, Y.~Zheng, M.~J. Black, O.~Hilliges, and A.~Geiger.
\newblock Snarf: Differentiable forward skinning for animating non-rigid neural
  implicit shapes.
\newblock In {\em International Conference on Computer Vision}, pages
  11594--11604, 2021.

\bibitem{cooks2022can}
E.~J. Cooks, K.~A. Duke, E.~Flood-Grady, M.~J. Vilaro, R.~Ghosh, N.~Parker,
  P.~Te, T.~J. George, B.~C. Lok, M.~Williams, et~al.
\newblock Can virtual human clinicians help close the gap in colorectal cancer
  screening for rural adults in the united states? the influence of rural
  identity on perceptions of virtual human clinicians.
\newblock {\em Preventive Medicine Reports}, 30:102034, 2022.

\bibitem{daniel2021perceptually}
B.~C. Daniel, R.~Marques, L.~Hoyet, J.~Pettr{\'e}, and J.~Blat.
\newblock A perceptually-validated metric for crowd trajectory quality
  evaluation.
\newblock {\em ACM on Computer Graphics and Interactive Techniques},
  4(3):1--18, 2021.

\bibitem{davydov2024cloaf}
A.~Davydov, M.~Engilberge, M.~Salzmann, and P.~Fua.
\newblock Cloaf: Collision-aware human flow.
\newblock {\em arXiv preprint arXiv:2403.09050}, 2024.

\bibitem{ding2020image}
K.~Ding, K.~Ma, S.~Wang, and E.~P. Simoncelli.
\newblock Image quality assessment: Unifying structure and texture similarity.
\newblock {\em Transactions on Pattern Analysis and Machine Intelligence},
  44(5):2567--2581, 2020.

\bibitem{fan2021point}
H.~Fan, Y.~Yang, and M.~Kankanhalli.
\newblock Point 4d transformer networks for spatio-temporal modeling in point
  cloud videos.
\newblock In {\em Conference on Computer Vision and Pattern Recognition}, pages
  14204--14213, 2021.

\bibitem{fan2022pstnet}
H.~Fan, X.~Yu, Y.~Ding, Y.~Yang, and M.~Kankanhalli.
\newblock Pstnet: Point spatio-temporal convolution on point cloud sequences.
\newblock {\em arXiv preprint arXiv:2205.13713}, 2022.

\bibitem{fernandes2021point}
D.~Fernandes, A.~Silva, R.~N{\'e}voa, C.~Sim{\~o}es, D.~Gonzalez, M.~Guevara,
  P.~Novais, J.~Monteiro, and P.~Melo-Pinto.
\newblock Point-cloud based 3d object detection and classification methods for
  self-driving applications: A survey and taxonomy.
\newblock {\em Information Fusion}, 68:161--191, 2021.

\bibitem{garland1997surface}
M.~Garland and P.~S. Heckbert.
\newblock Surface simplification using quadric error metrics.
\newblock In {\em Conference on Computer Graphics and Interactive Techniques},
  pages 209--216, 1997.

\bibitem{geijtenbeek2010evaluating}
T.~Geijtenbeek, A.~J. Van Den~Bogert, B.~J. Van~Basten, and A.~Egges.
\newblock Evaluating the physical realism of character animations using
  musculoskeletal models.
\newblock In {\em International Conference on Motion in Games}, pages 11--22.
  Springer, 2010.

\bibitem{ghildyal2022shift}
A.~Ghildyal and F.~Liu.
\newblock Shift-tolerant perceptual similarity metric.
\newblock In {\em European Conference on Computer Vision}, pages 91--107.
  Springer, 2022.

\bibitem{gulde2018smoothness}
P.~Gulde and J.~Hermsd{\"o}rfer.
\newblock Smoothness metrics in complex movement tasks.
\newblock {\em Frontiers in neurology}, 9:615, 2018.

\bibitem{herrera2020effect}
F.~Herrera, S.~Y. Oh, and J.~N. Bailenson.
\newblock Effect of behavioral realism on social interactions inside
  collaborative virtual environments.
\newblock {\em Presence}, 27(2):163--182, 2020.

\bibitem{hou2022perceptual}
Q.~Hou, A.~Ghildyal, and F.~Liu.
\newblock A perceptual quality metric for video frame interpolation.
\newblock In {\em European Conference on Computer Vision}, pages 234--253,
  2022.

\bibitem{huang2024humannorm}
X.~Huang, R.~Shao, Q.~Zhang, H.~Zhang, Y.~Feng, Y.~Liu, and Q.~Wang.
\newblock Humannorm: Learning normal diffusion model for high-quality and
  realistic 3d human generation.
\newblock In {\em Conference on Computer Vision and Pattern Recognition}, pages
  4568--4577, 2024.

\bibitem{jiang2022h4d}
B.~Jiang, Y.~Zhang, X.~Wei, X.~Xue, and Y.~Fu.
\newblock H4d: Human 4d modeling by learning neural compositional
  representation.
\newblock In {\em Conference on Computer Vision and Pattern Recognition}, pages
  19355--19365, 2022.

\bibitem{justice2022we}
J.~Justice, A.~Adkins, T.~Dong, and S.~J{\"o}rg.
\newblock Do we measure what we perceive? comparison of perceptual and computed
  differences between hand animations.
\newblock In {\em SIGGRAPH Posters}, pages 1--2. 2022.

\bibitem{Kovar2002}
L.~Kovar, J.~Schreiner, and M.~Gleicher.
\newblock Footskate cleanup for motion capture editing.
\newblock In {\em Symposium on Computer Animation}, page 97–104, 2002.

\bibitem{liu2019meteornet}
X.~Liu, M.~Yan, and J.~Bohg.
\newblock Meteornet: Deep learning on dynamic 3d point cloud sequences.
\newblock In {\em International Conference on Computer Vision}, pages
  9246--9255, 2019.

\bibitem{loper2023smpl}
M.~Loper, N.~Mahmood, J.~Romero, G.~Pons-Moll, and M.~J. Black.
\newblock {SMPL}: A skinned multi-person linear model.
\newblock In {\em Seminal Graphics Papers: Pushing the Boundaries, Volume 2},
  pages 851--866. 2023.

\bibitem{luo2012perceptual}
P.~Luo and M.~Neff.
\newblock A perceptual study of the relationship between posture and gesture
  for virtual characters.
\newblock In {\em Motion in Games}, pages 254--265, 2012.

\bibitem{mao2021generating}
W.~Mao, M.~Liu, and M.~Salzmann.
\newblock Generating smooth pose sequences for diverse human motion prediction.
\newblock In {\em International Conference on Computer Vision}, pages
  13309--13318, 2021.

\bibitem{marsot2022representing}
M.~Marsot, S.~Wuhrer, J.-S. Franco, and A.~H. Olivier.
\newblock Representing motion as a sequence of latent primitives, a flexible
  approach for human motion modelling.
\newblock {\em arXiv preprint arXiv:2206.13142}, 2022.

\bibitem{mihajlovic2022coap}
M.~Mihajlovic, S.~Saito, A.~Bansal, M.~Zollhoefer, and S.~Tang.
\newblock Coap: Compositional articulated occupancy of people.
\newblock In {\em Conference on Computer Vision and Pattern Recognition}, pages
  13201--13210, 2022.

\bibitem{mildenhall2021nerf}
B.~Mildenhall, P.~P. Srinivasan, M.~Tancik, J.~T. Barron, R.~Ramamoorthi, and
  R.~Ng.
\newblock Nerf: Representing scenes as neural radiance fields for view
  synthesis.
\newblock {\em Communications of the ACM}, 65(1):99--106, 2021.

\bibitem{min2024perceptual}
X.~Min, H.~Duan, W.~Sun, Y.~Zhu, and G.~Zhai.
\newblock Perceptual video quality assessment: A survey.
\newblock {\em arXiv preprint arXiv:2402.03413}, 2024.

\bibitem{nehme2023textured}
Y.~Nehm{\'e}, J.~Delanoy, F.~Dupont, J.-P. Farrugia, P.~Le~Callet, and
  G.~Lavou{\'e}.
\newblock Textured mesh quality assessment: Large-scale dataset and deep
  learning-based quality metric.
\newblock {\em Transactions on Graphics}, 42(3):1--20, 2023.

\bibitem{nehme2020visual}
Y.~Nehm{\'e}, F.~Dupont, J.-P. Farrugia, P.~Le~Callet, and G.~Lavou{\'e}.
\newblock Visual quality of 3d meshes with diffuse colors in virtual reality:
  Subjective and objective evaluation.
\newblock {\em Transactions on Visualization and Computer Graphics},
  27(3):2202--2219, 2020.

\bibitem{nehme2019comparison}
Y.~Nehm{\'e}, J.-P. Farrugia, F.~Dupont, P.~LeCallet, and G.~Lavou{\'e}.
\newblock Comparison of subjective methods, with and without explicit
  reference, for quality assessment of 3d graphics.
\newblock In {\em Symposium on Applied Perception}, pages 1--9, 2019.

\bibitem{patotskaya2023avoiding}
Y.~Patotskaya, L.~Hoyet, A.-H. Olivier, J.~Pettr{\'e}, and K.~Zibrek.
\newblock Avoiding virtual humans in a constrained environment: Exploration of
  novel behavioural measures.
\newblock {\em Computers \& Graphics}, 110:162--172, 2023.

\bibitem{pham2007formation}
Q.-C. Pham, H.~Hicheur, G.~Arechavaleta, J.-P. Laumond, and A.~Berthoz.
\newblock The formation of trajectories during goal-oriented locomotion in
  humans. ii. a maximum smoothness model.
\newblock {\em European Journal of Neuroscience}, 26(8):2391--2403, 2007.

\bibitem{pravzak2011perceptual}
M.~Pra{\v{z}}{\'a}k, L.~Hoyet, and C.~O'Sullivan.
\newblock Perceptual evaluation of footskate cleanup.
\newblock In {\em Symposium on Computer Animation}, pages 287--294, 2011.

\bibitem{reitsma2003perceptual}
P.~S. Reitsma and N.~S. Pollard.
\newblock Perceptual metrics for character animation: sensitivity to errors in
  ballistic motion.
\newblock In {\em SIGGRAPH}, pages 537--542. 2003.

\bibitem{rekik2024correspondence}
R.~Rekik, M.~Marsot, A.-H. Olivier, J.-S. Franco, and S.~Wuhrer.
\newblock Correspondence-free online human motion retargeting.
\newblock In {\em International Conference on 3D Vision}, pages 707--716, 2024.

\bibitem{rekik2024survey}
R.~Rekik, S.~Wuhrer, L.~Hoyet, K.~Zibrek, and A.-H. Olivier.
\newblock A survey on realistic virtual human animations: Definitions, features
  and evaluations.
\newblock In {\em Computer Graphics Forum}, 2024.

\bibitem{rempe2021humor}
D.~Rempe, T.~Birdal, A.~Hertzmann, J.~Yang, S.~Sridhar, and L.~Guibas.
\newblock Humor: 3d human motion model for robust pose estimation.
\newblock In {\em International Conference on Computer Vision}, pages
  11488--11499, 2021.

\bibitem{ren2005data}
L.~Ren, A.~Patrick, A.~A. Efros, J.~K. Hodgins, and J.~M. Rehg.
\newblock A data-driven approach to quantifying natural human motion.
\newblock {\em Transactions on Graphics}, 24(3):1090--1097, 2005.

\bibitem{rumman2017skin}
N.~A. Rumman and M.~Fratarcangeli.
\newblock Skin deformation methods for interactive character animation.
\newblock In {\em Computer Vision, Imaging and Computer Graphics Theory and
  Applications}, pages 153--174, 2017.

\bibitem{series2012methodology}
B.~Series.
\newblock Methodology for the subjective assessment of the quality of
  television pictures.
\newblock {\em Recommendation ITU-R BT}, 500(13), 2012.

\bibitem{shin2024wham}
S.~Shin, J.~Kim, E.~Halilaj, and M.~J. Black.
\newblock Wham: Reconstructing world-grounded humans with accurate 3d motion.
\newblock In {\em Conference on Computer Vision and Pattern Recognition}, pages
  2070--2080, 2024.

\bibitem{shiradkar2021virtual}
S.~Shiradkar, L.~Rabelo, F.~Alasim, and K.~Nagadi.
\newblock Virtual world as an interactive safety training platform.
\newblock {\em Information}, 12(6):219, 2021.

\bibitem{subramanyam2020comparing}
S.~Subramanyam, J.~Li, I.~Viola, and P.~Cesar.
\newblock Comparing the quality of highly realistic digital humans in 3dof and
  6dof: A volumetric video case study.
\newblock In {\em Conference on Virtual Reality and 3D User Interfaces}, pages
  127--136, 2020.

\bibitem{tseng2023edge}
J.~Tseng, R.~Castellon, and K.~Liu.
\newblock Edge: Editable dance generation from music.
\newblock In {\em Conference on Computer Vision and Pattern Recognition}, pages
  448--458, 2023.

\bibitem{TurnerWeta2017}
N.~Turner, M.~Reeves, J.~Letteri, D.~Lemmon, and D.~Barrett.
\newblock Weta digital vfx: War for the planet of the apes.
\newblock In {\em SIGGRAPH Computer Animation Festival}, page~27, 2017.

\bibitem{van2022between}
M.~van Kreveld, T.~Miltzow, T.~Ophelders, W.~Sonke, and J.~L. Vermeulen.
\newblock Between shapes, using the hausdorff distance.
\newblock {\em Computational Geometry}, 100:101817, 2022.

\bibitem{verkuyl2017virtual}
M.~Verkuyl, D.~Romaniuk, L.~Atack, and P.~Mastrilli.
\newblock Virtual gaming simulation for nursing education: An experiment.
\newblock {\em Clinical Simulation in Nursing}, 13(5):238--244, 2017.

\bibitem{villegas2018neural}
R.~Villegas, J.~Yang, D.~Ceylan, and H.~Lee.
\newblock Neural kinematic networks for unsupervised motion retargetting.
\newblock In {\em Conference on Computer Vision and Pattern Recognition}, pages
  8639--8648, 2018.

\bibitem{viviani1995minimum}
P.~Viviani and T.~Flash.
\newblock Minimum-jerk, two-thirds power law, and isochrony: converging
  approaches to movement planning.
\newblock {\em Journal of Experimental Psychology: Human Perception and
  Performance}, 21(1):32, 1995.

\bibitem{voas2023best}
J.~Voas, Y.~Wang, Q.~Huang, and R.~Mooney.
\newblock What is the best automated metric for text to motion generation?
\newblock In {\em SIGGRAPH Asia Conference Papers}, pages 1--11, 2023.

\bibitem{wang2024aligning}
H.~Wang, W.~Zhu, L.~Miao, Y.~Xu, F.~Gao, Q.~Tian, and Y.~Wang.
\newblock Aligning human motion generation with human perceptions.
\newblock {\em arXiv preprint arXiv:2407.02272}, 2024.

\bibitem{wang2020neural}
J.~Wang, C.~Wen, Y.~Fu, H.~Lin, T.~Zou, X.~Xue, and Y.~Zhang.
\newblock Neural pose transfer by spatially adaptive instance normalization.
\newblock In {\em Conference on Computer Vision and Pattern Recognition}, pages
  5831--5839, 2020.

\bibitem{wang2004image}
Z.~Wang, A.~C. Bovik, H.~R. Sheikh, and E.~P. Simoncelli.
\newblock Image quality assessment: from error visibility to structural
  similarity.
\newblock {\em Transactions on Image Processing}, 13(4):600--612, 2004.

\bibitem{wen2022point}
H.~Wen, Y.~Liu, J.~Huang, B.~Duan, and L.~Yi.
\newblock Point primitive transformer for long-term 4d point cloud video
  understanding.
\newblock In {\em European Conference on Computer Vision}, pages 19--35.
  Springer, 2022.

\bibitem{witte2022sports}
K.~Witte, M.~Droste, Y.~Ritter, P.~Emmermacher, S.~Masik, D.~B{\"u}rger, and
  K.~Petri.
\newblock Sports training in virtual reality to improve response behavior in
  karate kumite with transfer to real world.
\newblock {\em Frontiers in Virtual Reality}, 3:903021, 2022.

\bibitem{wolski2022geo}
K.~Wolski, L.~Trutoiu, Z.~Dong, Z.~Shen, K.~Mackenzie, and A.~Chapiro.
\newblock Geo-metric: A perceptual dataset of distortions on faces.
\newblock {\em Transactions on Graphics}, 41(6):1--13, 2022.

\bibitem{wu20244d}
G.~Wu, T.~Yi, J.~Fang, L.~Xie, X.~Zhang, W.~Wei, W.~Liu, Q.~Tian, and X.~Wang.
\newblock 4d gaussian splatting for real-time dynamic scene rendering.
\newblock In {\em Conference on Computer Vision and Pattern Recognition}, pages
  20310--20320, 2024.

\bibitem{wu2019novel}
Z.~Wu, W.~Jiang, H.~Luo, and L.~Cheng.
\newblock A novel self-intersection penalty term for statistical body shape
  models and its applications in 3d pose estimation.
\newblock {\em Applied Sciences}, 9(3):400, 2019.

\bibitem{xu2024finepose}
J.~Xu, Y.~Guo, and Y.~Peng.
\newblock Finepose: Fine-grained prompt-driven 3d human pose estimation via
  diffusion models.
\newblock In {\em Conference on Computer Vision and Pattern Recognition}, pages
  561--570, 2024.

\bibitem{xu2020rignet}
Z.~Xu, Y.~Zhou, E.~Kalogerakis, C.~Landreth, and K.~Singh.
\newblock Rignet: Neural rigging for articulated characters.
\newblock {\em arXiv preprint arXiv:2005.00559}, 2020.

\bibitem{yang2016estimation}
J.~Yang, J.-S. Franco, F.~H{\'e}troy-Wheeler, and S.~Wuhrer.
\newblock Estimation of human body shape in motion with wide clothing.
\newblock In {\em European Conference on Computer Vision}, pages 439--454,
  2016.

\bibitem{yang2023qpgesture}
S.~Yang, Z.~Wu, M.~Li, Z.~Zhang, L.~Hao, W.~Bao, and H.~Zhuang.
\newblock Qpgesture: Quantization-based and phase-guided motion matching for
  natural speech-driven gesture generation.
\newblock In {\em Conference on Computer Vision and Pattern Recognition}, pages
  2321--2330, 2023.

\bibitem{yu2023acr}
Z.~Yu, S.~Huang, C.~Fang, T.~P. Breckon, and J.~Wang.
\newblock Acr: Attention collaboration-based regressor for arbitrary two-hand
  reconstruction.
\newblock In {\em Conference on Computer Vision and Pattern Recognition}, pages
  12955--12964, 2023.

\bibitem{zhang2022avatargen}
J.~Zhang, Z.~Jiang, D.~Yang, H.~Xu, Y.~Shi, G.~Song, Z.~Xu, X.~Wang, and
  J.~Feng.
\newblock Avatargen: a 3d generative model for animatable human avatars.
\newblock In {\em European Conference on Computer Vision}, pages 668--685,
  2022.

\bibitem{zhang2023skinned}
J.~Zhang, J.~Weng, D.~Kang, F.~Zhao, S.~Huang, X.~Zhe, L.~Bao, Y.~Shan,
  J.~Wang, and Z.~Tu.
\newblock Skinned motion retargeting with residual perception of motion
  semantics \& geometry.
\newblock In {\em Conference on Computer Vision and Pattern Recognition}, pages
  13864--13872, 2023.

\bibitem{zhang2018unreasonable}
R.~Zhang, P.~Isola, A.~A. Efros, E.~Shechtman, and O.~Wang.
\newblock The unreasonable effectiveness of deep features as a perceptual
  metric.
\newblock In {\em Conference on Computer Vision and Pattern Recognition}, pages
  586--595, 2018.

\bibitem{zhao2021m3d}
F.~Zhao, Z.~Xie, M.~Kampffmeyer, H.~Dong, S.~Han, T.~Zheng, T.~Zhang, and
  X.~Liang.
\newblock M3d-vton: A monocular-to-3d virtual try-on network.
\newblock In {\em International Conference on Computer Vision}, pages
  13239--13249, 2021.

\bibitem{zhou2024simple}
Z.~Zhou, S.~Zhou, Z.~Lv, M.~Zou, Y.~Tang, and J.~Liang.
\newblock A simple baseline for efficient hand mesh reconstruction.
\newblock In {\em Conference on Computer Vision and Pattern Recognition}, pages
  1367--1376, 2024.

\bibitem{zhu2022deepdc}
H.~Zhu, B.~Chen, L.~Zhu, S.~Wang, and W.~Lin.
\newblock Deepdc: Deep distance correlation as a perceptual image quality
  evaluator.
\newblock {\em arXiv e-prints}, pages arXiv--2211, 2022.

\bibitem{zibrek2019photorealism}
K.~Zibrek, S.~Martin, and R.~McDonnell.
\newblock Is photorealism important for perception of expressive virtual humans
  in virtual reality?
\newblock {\em Transactions on Applied Perception}, 16(3):1--19, 2019.

\bibitem{zins2023multi}
P.~Zins, Y.~Xu, E.~Boyer, S.~Wuhrer, and T.~Tung.
\newblock Multi-view reconstruction using signed ray distance functions (srdf).
\newblock In {\em Conference on Computer Vision and Pattern Recognition}, pages
  16696--16706, 2023.

\bibitem{zou2020reducing}
Y.~Zou, J.~Yang, D.~Ceylan, J.~Zhang, F.~Perazzi, and J.-B. Huang.
\newblock Reducing footskate in human motion reconstruction with ground contact
  constraints.
\newblock In {\em Winter Conference on Applications of Computer Vision}, pages
  459--468, 2020.

\end{thebibliography}
